\input harvmac 
\input epsf.tex
\def\IN{\relax{\rm I\kern-.18em N}} 
\def\IR{
\relax{\rm I\kern-.18em R}} \font\cmss=cmss10 
\font\cmsss=cmss10 at 7pt \def\IZ{\relax\ifmmode\mathchoice 
{\hbox{\cmss Z\kern-.4em Z}}{\hbox{\cmss Z\kern-.4em Z}} 
{\lower.9pt\hbox{\cmsss Z\kern-.4em Z}} {\lower1.2pt\hbox{
\cmsss Z\kern-.4em Z}}
\else{\cmss Z\kern-.4em Z}\fi}

\overfullrule=0mm
\def\file#1{#1}
\newcount\figno \figno=0
\newcount\figtotno      
\figtotno=0
\newdimen\captionindent 
\captionindent=1cm 
\def\figbox#1#2{\epsfxsize=#1\vcenter{
\epsfbox{\file{#2}}}} 
\newcount\figno
\figno=0
\def\fig#1#2#3{ 
\par\begingroup\parindent=0pt
\leftskip=1cm\rightskip=1cm\parindent =0pt 
\baselineskip=11pt
\global\advance\figno by 1
\midinsert
\epsfxsize=#3
\centerline{\epsfbox{#2}}
\vskip 12pt
{\bf Fig. \the\figno:} #1\par
\endinsert\endgroup\par
}
\def\figlabel#1{\xdef#1{\the\figno}} 
\def\encadremath#1{\vbox{\hrule\hbox{\vrule\kern8pt 
\vbox{\kern8pt \hbox{$\displaystyle #1$}\kern8pt} 
\kern8pt\vrule}\hrule}} \def\enca#1{\vbox{\hrule\hbox{
\vrule\kern8pt\vbox{\kern8pt \hbox{$\displaystyle #1$}
\kern8pt} \kern8pt\vrule}\hrule}}
\def\tvi{\vrule height 12pt depth 6pt width 0pt} 
\def\tv{\tvi\vrule}

\def\IR{\relax{\rm I\kern-.18em R}}
\font\cmss=cmss10 \font\cmsss=cmss10 at 7pt 
\def\IZ{\relax\ifmmode\mathchoice
{\hbox{\cmss Z\kern-.4em Z}}{\hbox{\cmss Z\kern-.4em Z}} 
{\lower.9pt\hbox{\cmsss Z\kern-.4em Z}}
{\lower1.2pt\hbox{\cmsss Z\kern-.4em Z}} 
\else{\cmss Z\kern-.4em Z}\fi} \def\buildrel#1\under#2{ 
\mathrel{\mathop{\kern0pt #2}\limits_{#1}}}
\def\b{\beta}
\def\leg#1{ 
\par\begingroup\parindent=0pt
\leftskip=1cm\rightskip=1cm\parindent =0pt 
\baselineskip=11pt
#1 \par \endgroup\par}

\Title{UNC-CH-MATH-97/4}
{{\vbox {
\bigskip
\centerline{SU(N) Meander Determinants}
}}}
\bigskip
\centerline{P. Di Francesco\footnote*{e-mail: philippe@math.unc.edu},}
\bigskip
\centerline{Department of Mathematics,} 
\centerline{\it University of North Carolina at Chapel Hill,} 
\centerline{\it  CHAPEL HILL, N.C. 27599-3250, U.S.A.} 
\centerline{and}
\centerline{ \it Service de Physique Th\'eorique,} 
\centerline{C.E.A. Saclay,}
\centerline{ \it F-91191 Gif sur Yvette Cedex, France} 
\vskip .5in
\noindent We propose a generalization of meanders, i.e., 
configurations of non-selfintersecting loops crossing a line
through a given number of points, to $SU(N)$. This uses the 
reformulation of meanders as pairs of reduced elements of
the Temperley-Lieb algebra,
a $SU(2)$-related quotient of the Hecke algebra, with a natural
generalization to $SU(N)$.
We also derive explicit
formulas for $SU(N)$ meander determinants, defined as the Gram
determinants of the corresponding bases of the Hecke algebra.

\Date{02/97}
\nref\TOU{J. Touchard, 
{\it Contributions \`a l'\'etude du probl\`eme des timbres poste}, 
Canad. J. Math. {\bf 2} (1950) 385-398.}
\nref\LUN{W. Lunnon, {\it A map--folding problem}, 
Math. of Computation {\bf 22} 
(1968) 193-199.}
\nref\HMRT{K. Hoffman, K. Mehlhorn, P. Rosenstiehl and
R. Tarjan, {\it Sorting Jordan sequences in linear time 
using level-linked 
search trees}, Information and Control {\bf 68} (1986) 170-184.} 
\nref\ARNO{V. Arnold, {\it The branched covering of $CP_2 \to S_4$, 
hyperbolicity and projective topology}, 
Siberian Math. Jour. {\bf 29} (1988) 717-726.} 
\nref\LZ{S. Lando and A. Zvonkin, 
{\it Plane and Projective Meanders}, Theor. Comp. 
Science {\bf 117} (1993) 227-241, 
and {\it Meanders}, Selecta Math. Sov. {\bf 11} (1992) 117-144.} 
\nref\KOSMO{K.H. Ko, L. Smolinsky, {\it A combinatorial matrix 
in $3$-manifold theory}, 
Pacific. J. Math {\bf 149} (1991) 319-336.} 
\nref\DGG{P. Di Francesco, O. Golinelli 
and E. Guitter, {\it Meander,
folding and arch statistics}, 
to appear in Math. and Comp. 
Modelling (1997), hep-th/950630.}
\nref\DGGB{P. Di Francesco, O. Golinelli and E. Guitter, {\it 
Meanders and the Temperley-Lieb algebra}, to appear in Comm. Math. Phys.
(1997), hep-th/9602025.}
\nref\TLA{H. Temperley and E. Lieb, 
{\it Relations between the percolation and coloring problem and 
other graph-theoretical problems associated with regular planar 
lattices: some exact results for the percolation problem}, 
Proc. Roy. Soc. {\bf A322} 
(1971) 251-280.}
\nref\MARTIN{P. Martin, 
{\it Potts models and related problems in statistical mechanics}, 
World Scientific (1991).}
\nref\MOI{P. Di Francesco, {\it Meander determinants}, preprint
UNC-CH-MATH-96/1 (1996) hep-th/9612026.}
\nref\RESHE{N. Reshetikhin, {\it Quantized universal 
enveloping algebras, the Yang-Baxter equation and invariants of links 
1 and 2}, LOMI preprints E-4-87 and E-17-87 (1988).}
\nref\JM{
M. Jimbo, T. Miwa and M. Okado, {\it An $A_{n-1}^{(1)}$ family
of solvable lattice models},
Mod. Phys. Lett. {\bf B1} (1987) 73-79, and {\it Solvable lattice models
related to the vector representation of classical simple Lie algebras},
Comm. Math. Phys. {\bf 116} (1988) 507-525.}
\nref\DJ{R. Dipper and G. James, {\it Representations of Hecke algebras of
general linear groups}, Proc. London Math. Soc. {\bf 52} (1986) 20-52;
{\it Blocks and idempotents of Hecke
algebras of general linear groups}, Proc. London Math. Soc.
{\bf 54} (1987) 57-82.}
\nref\CHE{I. Cherednik,{\it A new interpretation of Gelfand-Tzeitlin
bases}, Duke Math. Jour. {\bf 54} (1987) 563-577.}
\nref\MUR{G. Murphy, {\it On the representation theory of the symmetric
groups and associated Hecke algebras}, Jour. Algebra {\bf 152} (1992)
492-513.}
\nref\LRD{A. Kuniba and T. Nakanishi, {\it Level-rank duality
in fusion RSOS models}, in Proc. of the international colloquium on
modern quantum field theory, Bombay, Eds S. Das et al., World Scientific
(1991).}
\nref\DGGT{P. Di Francesco, O. Golinelli and E.Guitter, 
{\it Meanders: a direct enumeration approach}, Nucl. Phys. 
{\bf B482[FS]} (1997) 497-535.}

\newsec{Introduction}

In this paper we propose various generalizations of the concept of
meander \TOU\ \LUN\ \HMRT\ \ARNO\ \LZ. 
The original meander problem  
consists in counting the number 
$M_n$ of meanders of order $n$, i.e. of 
topologically inequivalent configurations
of a closed non-self-intersecting
loop crossing an infinite line through $2n$ points.  
One can also define the corresponding multi-component meander
problem, by demanding that the loop be replaced by a given number 
of non-intersecting loops (connected
components).
The meander problem probably first arose in the work of 
Poincar\'e about differential geometry, 
then reemerged in various contexts, such as 
the classification 
of 3-manifolds \KOSMO, or the physics of compact polymer folding
\DGG.  

In the present paper, we extend  
the purely algebraic approach advocated in \DGGB, 
which relates multi-component meanders to {\it pairs} of reduced
elements of the Temperley-Lieb algebra \TLA\ 
(see also P. Martin's book
\MARTIN\ for an elementary introduction), or ideals thereof. 
The idea is to define generalized multi-component meanders
as {\it pairs} of reduced elements of the $SU(N)$ quotients
of the Hecke algebra \RESHE\
which generalize the Temperley-Lieb $SU(2)$
quotient, or of ideals thereof.
The notion of ``component" for generalized meanders 
still awaits a good combinatorial
interpretation. We trade it in the present
approach for a piece of information 
on any given generalized meander, provided by the 
Markov trace of the
corresponding product of reduced elements.
Given a reduced basis of the above Hecke algebra quotients or ideals, 
this information is summarized by the Gram matrix of the
basis.
The aim of this work is to compute explicitly the ``meander determinants"
namely the determinants of these Gram matrices.

\par
\medskip
The paper is organized as follows. 
In Sect.2, we recall basic definitions and summarize the results
obtained in \DGGB\ and \MOI\ for the $SU(2)$ meander determinant,
in the form of an explicit determinantal formula. We also present
the $SU(N)$ quotients of the Hecke algebra, generalizing the 
Temperley-Lieb algebra. 
In Sect.3, we focus our attention on the $SU(3)$ case.
We are led to the natural definition of $SU(3)$ meanders, as
pairs of elements of
the reduced basis of a certain ideal ${\cal I}_{3n}^{(3)}(\beta)$
of the $SU(3)$ quotient $H_{3n}^{(3)}(\beta)$ of the Hecke algebra. 
This basis is labelled by closed paths
of length $3n$ on the Weyl chamber of $sl(3)$, the $SU(3)$ walk
diagrams. 
We then compute the corresponding Gram determinant, by direct
orthogonalization of the basis. We obtain an explicit
formula for the $SU(3)$ meander determinant. 
This result is generalized to $SU(N)$ in Sect.4, where we also
establish a duality relation between the ideals ${\cal I}_{Nk}^{(N)}(\beta)$
and ${\cal I}_{Nk}^{(k)}(\beta)$, relating the
$SU(N)$ and $SU(k)$ meander determinants.
In Sect.5, we derive a determinant formula for the Gram matrix of
a reduced basis of the whole $SU(N)$ quotient $H_{n}^{(N)}(\beta)$
of the Hecke algebra. This coincides with the meander determinant
only in the $SU(2)$ case, and suggests another possible generalization
of meanders. 
We gather a few concluding remarks in Sect.6.

\par
\newsec{Meanders and SU(2)}
\par
\subsec{Definitions}
\par
The meander problem of order $2n$
is that of enumerating the topologically
inequivalent configurations of a planar non-intersecting closed road
(loop) crossing a river (line) through $2n$ distinct bridges.
A meander is therefore represented as a non-self-intersecting
loop crossing a line  through $2n$ distinct points. 
The line cuts the meander into an upper and a lower part, which are both
made of $n$ non-intersecting arches (pieces of the loop) connecting
the $2n$ bridges by pairs. Such an upper (or lower) configuration of
a meander is called an arch configuration of order $2n$.
The set of arch configurations of order $2n$, $A_{2n}$, has cardinal
equal to the Catalan number
\eqn\catal{c_n~=~{(2n)!\over  (n+1)!n!} } 
readily proved by induction.

$$\vbox{\font\bidon=cmr8 \def\bidondon{\bidon} \bidondon
\offinterlineskip
\halign{\tv \quad # \tv &
\hfill \ # & \hfill # & \hfill # &  \hfill #
& \hfill # &  \hfill # &  \hfill #
&  \hfill # &  \hfill # &  \hfill #
\tv \cr
\noalign{\hrule}
\tvi  $n$& 1 \hfill & 2 \hfill
& 3 \hfill & 4 \hfill & 5 \hfill
& 6 \hfill & 7 \hfill & 8 \hfill
& 9 \hfill & 10 \hfill 
\cr
\noalign{\hrule}
\tvi $c_n$ & 1 & 2 & 5 & 14 & 42 & 132 & 429 & 1430 & 4862 &
16796  \cr
\noalign{\hrule}
}}$$
\leg{{\bf Table I:} The Catalan numbers for $n=1,2,...,10$.}

A multi-component meander of order $2n$
is the superposition of two
arbitrary upper and lower arch configurations $a$, $b\in A_{2n}$.
This results a priori in a configuration of $k$ different
non-intersecting roads crossing the river through a total of $2n$
bridges: $k$ is called the number of connected components of the
meander,
also denoted by $k=\kappa(a,b)$. 

We choose to adopt an alternative description of 
meanders in terms of $SU(2)$ walk diagrams as follows.
A $SU(2)$ walk diagram of order $2n$
is a closed path of length $2n$ on the semi-infinite line
$\{1,2,3,...\}$ identified with the Weyl chamber of the $sl(2)$ Lie
algebra. More precisely, a walk diagram is a sequence
$\{ h(i), \ i=0,1,2,...,2n\}$ of positive integer ``heights", such that
\fig{A sample walk diagram of order $10$.}{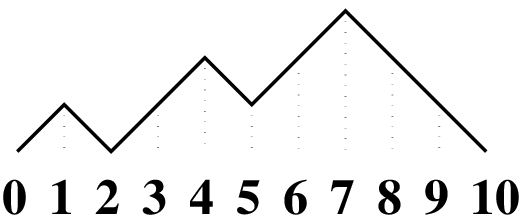}{5.cm}
\figlabel\picto
\eqn\conswa{ h(i+1)-h(i) \in \{ 1, -1\} \qquad h(0)=h(2n)=1 }
A pictorial representation for a walk diagram is presented in
Fig.\picto: it consists of the graph of the 
corresponding function $i \to h(i)$, whose points are joined by
consecutive segments. We denote by $W_{2n}^{(2)}$ the set of 
walk diagrams\foot{Here and in the following,
the superscript $(2)$ stands for $SU(2)$.} 
of order $2n$.

The walk diagrams of order $2n$
are in one-to-one correspondence with the 
arch configurations of order $2n$. Starting from an arch configuration
of order $2n$ let us label by $0$, $1$, $2$, ..., $2n$ respectively
the portions of river to the left of the leftmost bridge, 
between the first and
second, ...,  to the right of the rightmost bridge along the river. 
We define the map $i \to h(i)$ by assigning to the portion of river
labelled $i$ the number $h(i)$ of arches passing above it, plus one\foot{
This is slightly different from the 
conventions of refs.\DGGB\ and \MOI, in which $h(i)\geq 0$ rather
than $1$. Our present choice is motivated by the form of the forthcoming 
$SU(N)$ and Hecke generalizations.}.

The constraints \conswa\ are satisfied by $h$ hence we have
constructed a walk diagram for each arch configuration; the process is
clearly bijective, as an arch configuration is entirely determined
by the numbers $h(i+1)-h(i)$,  with the value $+1$ if an arch 
originates from the left bridge of the portion $i$ of river
and passes above it,
and the value $-1$ if an arch terminates at the left bridge of the
portion $i$ of river (and therefore does not pass over it). 

A multi-component meander of order $2n$
is  therefore equivalently given by a couple
$(a,b)$ of walk diagrams of order $2n$, and we still denote by
$\kappa(a,b)$ its number of connected components of road.

\subsec{Temperley-Lieb algebra}
\par
The link between the above arch configuration and walk diagram pictures
is provided by the Temperley-Lieb algebra, as well as a direct
interpretation of the quantity $\kappa(a,b)$, for $a,b \in W_{2n}^{(2)}$.

The Temperley-Lieb algebra $TL_n(\beta)$ is defined by generators
$1,e_1,e_2,...,e_{n-1}$ and relations
\eqn\tla{ \eqalign{
e_i^2~&=~ \beta\ e_i \ \ \ \ {\rm for} \ i=1,2,...,n-1 \cr
e_i e_j ~&=~ e_j e_i \ \ \ \ {\rm for}\ |i-j|>1 \cr
e_i e_{i\pm 1} e_i ~&=~ e_i \ \ \ \ {\rm for} \ i=1,2,...,n-1 \cr}}
An element of this algebra is said to be reduced if it is written
as a product of generators, with a minimal number of them 
(``reduction" is achieved by repeated use of the relations \tla).

For reasons which will become clear later, we will work with
a certain left ideal of the Temperley-Lieb algebra $TL_{2n}(\beta)$,
which is however isomorphic to $TL_n(q)$.
We denote by ${\cal I}_{2n}^{(2)}(\beta)$ the left ideal generated by the
element $e_1 e_3 e_5 ...e_{2n-1}$ of $TL_{2n}(\beta)$.

There is a one-to-one correspondence between the reduced elements 
of the ideal ${\cal I}_{2n}^{(2)}(\beta)$ and the walk diagrams of
order $2n$. To best see this, let us first reconsider the walk diagrams
of order $2n$. We start from the ``fundamental" walk diagram
$a_0^{(2)}\in W_{2n}^{(2)}$, such that
\eqn\funda{ h(1)=h(3)=...=h(2n-1)=2 \ \ {\rm and}\ \
h(0)=h(2)=...=h(2n)=1}
This is the walk with the smallest values of the height $h(i)$.
Now any other walk diagram of order $2n$ may be constructed by
successive ``box additions" on $a_0^{(2)}$. By box addition on a
walk diagram $a$ at position $i$, 
which we denote by $a+\diamond_i$, we mean
the following transformation.
For the box addition to be possible, $a$ must have a minimum at the
position $i$, namely $h(i+1)=h(i-1)=h(i)+1$. The box addition then
simply amounts to transform this minimum into a maximum, namely
change $h(i) \to h(i)+2$, without altering the other values of $h$.
By successive box additions on $a_0^{(2)}$, it is easy
to describe all the set of walk diagrams of order $2n$.  
Note that a given walk diagram may be obtained by distinct sequences of
box additions on $a_0^{(2)}$, 
but all of them will consist of the same total
number of box additions.
We are now in position to construct a map $\varphi$
from $W_{2n}^{(2)}$ to a basis
of reduced elements of ${\cal I}_{2n}^{(2)}(\beta)$.
We start with  
\eqn\stamap{ \varphi(a_0^{(2)})~=~ e_1 e_3 ... e_{2n-1} }
and proceed recursively, using box additions, by setting
\eqn\recuset{ \varphi(a+\diamond_i)~=~ e_i\  \varphi(a) }
The map is well-defined, as two distinct sequences of box additions
leading to the same walk diagram correspond to different
products of the same commuting $e_i$'s (at each step, if two distinct
box additions are possible, they take place at positions $i$ and $j$
with $|j-i|>1$, hence the corresponding $e_i$ and $e_j$ commute,
due to \tla).
It exhausts all the reduced elements of ${\cal I}_{2n}^{(2)}(\beta)$,
which has the dimension $c_n$ \catal\ as a vector space.

A meander is therefore equivalently given by a pair of reduced
elements of ${\cal I}_{2n}^{(2)}(\beta)$.
The Temperley-Lieb algebra $TL_n(\beta)$
is endowed with a natural scalar product
attached to the Markov trace, denoted by ${\rm Tr}$.
The latter is defined by the normalization
${\rm Tr}(1)=\beta^{n}$, and the Markov
property that for any element $E(e_1,e_2,...,e_{j-1})$ involving 
only $e_i$'s with $i<j$, we have
\eqn\marko{ {\rm Tr} \big(E(e_1,e_2,...,e_{j-1}) e_j\big)
~=~ {\eta} {\rm Tr} \big( E(e_1,e_2,...,e_{j-1})\big)}
The standard choice for $TL_n(\beta)$ for the constant $\eta$ is
\eqn\coicet{ \eta~=~{1\over \beta} }
The trace extends linearly to any element of $TL_n(\beta)$.
We also need to define the transposed $e^t$ of 
an element $e\in TL_n(\beta)$,
as $1^t=1$, $e_i^t=e_i$ for $i=1,2,...,n$, and $(ef)^t=f^t e^t$ for
any two elements $e,f\in TL_n(\beta)$; again, the definition extends to
any element by linearity. 
This leads to the scalar product
\eqn\scapro{ (e,f)~=~ {\rm Tr} ( e f^t) }
Remarkably, when restricted to the ideal ${\cal I}_{2n}^{(2)}(\beta)$,
and when expressed between two reduced elements say $\varphi(a)$
and $\varphi(b)$, $a$, $b$ two walk diagrams of order $2n$, this 
scalar product reads
\eqn\scapred{ (\varphi(a),\varphi(b))~=~ \beta^{\kappa(a,b)+n} }
thus making the contact with our initial road/river picture of
meanders.
Defining the normalized reduced
basis elements $(a)_1=\beta^{-n/2} \varphi(a)$ (this basis is referred to
as basis 1 in the following),
we have
\eqn\basone{ \big( (a)_1,(b)_1 \big)~=~ \beta^{\kappa(a,b)} }

\subsec{Meander determinant}
\par
The meander determinant $\Delta_{2n}^{(2)}(\beta)$
is defined as the determinant of the Gram matrix
of the basis 1 above, namely the $c_n \times c_n$
matrix ${\cal G}_{2n}^{(2)}(\beta)$ with entries 
\eqn\grammat{\big[{\cal G}_{2n}^{(2)}(\beta)\big]_{a,b}
~=~\beta^{\kappa(a,b)} } 
which therefore carries 
information about the multi-component meanders. 

In \DGGB\ \MOI, we have derived an exact formula 
for $\Delta_{2n}^{(2)}(\beta)$
based on the explicit Gram-Schmidt orthogonalization of the
matrix ${\cal G}_{2n}^{(2)}(\beta)$.
The formula reads
\eqn\detforme{ \Delta_{2n}^{(2)}(\beta)~=~\prod_{m=1}^n 
\big[U_m(\beta)\big]^{a_{m,n}^{(2)}} }
where $U_m(\beta)$ are the Chebishev polynomials of the first kind,
with 
\eqn\ceby{U_m(2 \cos \theta)~=~{\sin (m+1)\theta \over \sin \theta}}
and 
\eqn\vala{a_{m,n}^{(2)}~=~C_{2m+1}^{(2n)}-C_{2m+3}^{(2n)} }
where
$C_{2m+1}^{(2n)}$ counts the number of paths of length $2n$
on the half-line, starting from the origin ($h(0)=1$) 
and terminating at height $2m+1$ ($h(2n)=2m+1$), easily
computed as
\eqn\pathnum{ C_{2m+1}^{(2n)}~=~ {2n \choose n-m} -{2n \choose n-m-1}}
and in particular $C_1^{(2n)}=c_n$ of \catal.

$$\vbox{\font\bidon=cmr8 \def\bidondon{\bidon} \bidondon
\offinterlineskip
\halign{\tv \quad # \tv &
\hfill \ # & \hfill # & \hfill # &  \hfill #
& \hfill # &  \hfill # &  \hfill #
&  \hfill # 
&  \hfill # 
&  \hfill # 
\tv \cr
\noalign{\hrule}
\tvi  $\scriptstyle m\backslash n$& 1 \hfill & 2 \hfill
& 3 \hfill & 4 \hfill & 5 \hfill
& 6 \hfill & 7 \hfill & 8 \hfill & 9 \hfill & 10 \hfill\cr
\noalign{\hrule}
\tvi 1 & 1 & 2 & 4 & 8 & 15 & 22  & 0 & -208  & -1326 & -6460\cr
\tvi 2 &  & 1 & 4 & 13 & 40 & 121 & 364  & 1092 & 3264 & 9690  \cr
\tvi 3 &  &  & 1 & 6 & 26 & 100 & 364 & 1288 &  4488 & 15504 \cr
\tvi 4 &  &  &  & 1 & 8 & 43 & 196  & 820 & 3264 & 12597  \cr
\tvi 5 &  &  &  &   & 1  & 10  & 64 & 336 & 1581 & 6954\cr
\tvi 6 &  &  &  &   &   & 1  & 12  & 89 & 528 & 2755 \cr
\tvi 7 &  &  &  &   &   &    & 1 & 14 & 118 & 780 \cr
\tvi 8 &  &  &  &   &   &    &   &  1 & 16 & 151  \cr
\tvi 9 &  &  &  &   &   &    &   &   & 1 & 18  \cr
\tvi 10 &  &  &  &   &   &    &   &   & & 1  \cr
\noalign{\hrule}
}}$$
\leg{{\bf Table II:} The powers $a_{m,n}^{(2)}$ of $U_m$ in
the meander determinant of order $2n$, $\Delta_{2n}^{(2)}(\beta)$,
for $n=1,2,...,10$.}

\subsec{Generalizations}
\par

The remainder of this paper consists of various generalizations of this
determinant formula. The above discussion is strongly related
to the $sl(2)$ Lie algebra. Apart from the fact that
we considered paths on the Weyl chamber of $sl(2)$ (the half-line),
the Temperley-Lieb algebra is known to be a certain quotient of the
Hecke algebra $H_n(\beta)$. 
The latter is defined by generators $1$, $e_1$, $e_2$,...
$e_{n-1}$ and relations
\eqn\hec{\eqalign{  e_i^2~&=~ \beta e_i \ \ \ \ {\rm for} \
i=1,2,...,n-1 \cr
e_i e_j ~&=~ e_j e_i \ \ \ \ {\rm for} \ |i-j|>1 \cr
e_i e_{i+1} e_i -e_i ~&=~ e_{i+1} e_i e_{i+1} -e_{i+1} \ \ {\rm for}\ 
i=1,2,..,n-2 \cr}}
This algebra is usually defined through the generators
\eqn\genet{T_i~=~q^{1/2} (q^{1/2}-e_i)} 
where $\beta=q^{1/2}+q^{-1/2}$, as
a deformation of the symmetric group algebra (in particular, 
the three-term relation reads simply $T_iT_{i+1}T_i=
T_{i+1}T_iT_{i+1}$).
In terms of these latter generators, the quantites $e_i e_{i+1} e_i-e_i$,
by which we have to quotient the algebra to recover $TL_n(\beta)$
(see \tla),
are simply the generalized Young
antisymmetrizers of order $3$, namely
\eqn\antisym{
A(T_i,T_{i+1})~=~ 1-q^{-1}T_i- q^{-1}T_{i+1} + 
q^{-2}T_i T_{i+1}+q^{-2}T_{i+1} T_i- q^{-3}T_i T_ {i+1}T_i}
easily reexpressed in terms of the $e_i$'s as
\eqn\youngant{
Y(e_i,e_{i+1})~=~q^{3/2}A(T_i,T_{i+1})~=~e_i e_{i+1} e_i -e_i }
Requiring the vanishing of \antisym\ bears a strong analogy 
with the $SU(2)$ representations (allowing only for Young
tableaux with at most two lines), which can actually be 
made very precise,
and we will return to it in later sections\foot{The special 
$SU(N)$ quotients
of the Hecke algebra we will consider are also known as the commutants
of the quantum enveloping algebras $U_q(sl(N))$ \RESHE, 
and appear in the definition of the $A_{N-1}$ RSOS models of \JM.}.

For the moment, we will content ourselves with the natural
generalizations (to $SU(N)$)
of the Temperley-Lieb algebra by performing quotients
of the Hecke algebra by the generalized Young antisymmetrizer of order
$N+1$, $A(T_1,T_2,...,T_N)\equiv A(e_1,...,e_{N})$
\eqn\younga{ A(e_1,e_{2},...,e_{N})~=~ \sum_{w \in S_{N+1}}
(-q)^{-l(w)} T_w }
and its shifted versions under $e_j\to e_{j+i-1}$, for $j=1,2,...,N-1$.
In \younga,
the sum extends over all the permutations of $N+1$ objects,
$l(w)$ is the length of the permutation (the number of factors in any
minimal expression of $w$ as a product over transpositions of neighbors
$(i,i+1)$), and $T_w=T_{i_1} T_{i_2} ... T_{i_{l(w)}}$ if
$w=(i_1,i_1+1)(i_2,i_2+1)...(i_{l(w)},i_{l(w)}+1)$ (note that
this expression is independent of the particular minimal decomposition
of $w$,
thanks to the relation $T_i T_{i+1} T_i=T_{i+1}T_iT_{i+1}$). 
We will denote by $H_n^{(N)}(\beta)$ the corresponding 
$SU(N)$ quotient of the Hecke algebra. In particular, we have 
$TL_n(\beta)=H_n^{(2)}(\beta)$.

In terms of the Murphy operators \DJ\ \MUR, defined as
\eqn\murop{ L_m= q^{-1} T_{m-1} +q^{-2} T_{m-2} T_{m-1}T_{m-2}+...
+q^{-m+1} T_1 T_2 ...T_{m-2} T_{m-1} T_{m-2} ... T_2 T_1}
for $m\geq 2$, $L_1=0$,
it is possible to write compact expressions for the Young
antisymmetrizers of order $N$:
\eqn\compyou{ A(e_1,e_2,...,e_{N-1})~=~ \prod_{m=2}^{N} (1-L_m) }
In the following, we will use the various following versions 
of the Young antisymmetrizer of order $N$, 
which are all proportional to $A$
\younga:
\eqn\yonot{\eqalign{ y(e_1,...,e_{N-1})~&=~\prod_{m=2}^N
{1-L_m\over 1+q^{-1}+...+q^{-m+1}} \cr
E(e_1,...,e_{N-1})~&=~\prod_{m=2}^N q^{m-1 \over 2} (1-L_m)\cr}}
The antisymmetrizer $y(e_1,...,e_{N-1})$ is idempotent, $y^2=y$.
As mentioned before,
the argument $(e_1,...,e_{N-1})$ of $A,y,E$ may be shifted
into $(e_i,...,e_{i+N-2})$, and the corresponding functions
may be expressed
through analogous products, by performing the same shifts in $L_m$.
Finally, we will also use the following version of the 
Young antisymmetrizer, which has the advantage of being simply 
expressed in terms of the $e_i$'s, through a recursion (see \MARTIN),
starting with 
$Y(e_i)=e_i$, and
\eqn\recyou{Y(e_i,e_{i+1},...,e_{i+p})~=~
Y(e_i,...,e_{i+p-1})(e_{i+p}-\mu_{p}) Y(e_i,...,e_{i+p-1})}
for all $i,p\geq 1$, where we have introduced the quantities
\eqn\nota{ \mu_p~\equiv~\mu_p(\beta)~=~ {U_{p-1}(\beta) \over
U_p(\beta)} }
in terms of the Chebyshev polynomials \ceby, for all $p\geq 1$.
In particular, we have 
\eqn\ytwo{ Y(e_i,e_{i+1})~=~ 
e_i( e_{i+1}-\mu_1) e_i~=~e_ie_{i+1}e_i-e_i}
as $\mu_1=\beta^{-1}$ and $e_i^2=\beta e_i$.
The three antisymmetrizers $y,E,Y$ are proportional to $A$. In particular
we have 
\eqn\propY{\eqalign{
y(e_i,...,e_{i+N-2})~&=~\alpha_N E(e_i,...,e_{i+N-2})\cr
y(e_i,...,e_{i+N-2})~&=~\gamma_N Y(e_i,...,e_{i+N-2})\cr }}
where we have introduced the proportionality constants
\eqn\procons{\eqalign{
\alpha_N~&=~\prod_{i=1}^{N-1} (\mu_i)^{N-i} \cr
\gamma_N~&=~\prod_{i=1}^{N-1} (\mu_i)^{2^{N-i-1}}\cr}}
with $\alpha_{N+1}/\alpha_N=\mu_1\mu_2...\mu_N$, and
$\gamma_{N+1}/(\gamma_N)^2=\mu_N$.
The second relation of \propY\ is proved by induction on $N$,
by first showing that
\eqn\commu{ (1-L_{N+1})(e_N-q^{-1/2})~=~(e_N-q^{1/2})(1-L_N)+q^{-1/2}}
(also valid for any shift of the $e$'s), and finally deducing
that
\eqn\recy{ y(e_i,...,e_{i+N-1})~=~\mu_N y(e_i,...,e_{i+N-2})
(e_{i+N-1}-\mu_{N-1}) y(e_i,...,e_{i+N-2})}
As $y$ is idempotent, we also have the relation
\eqn\idmpo{
Y(e_i,...,e_{i+N-2})^2~=~\gamma_N^{-1} Y(e_i,...,e_{i+N-2})}

In the following, we suggest a generalization of meanders into pairs 
of $SU(N)$ walk
diagrams (see definitions below), and the meander determinant will
be generalized into the Gram determinant of the basis of some ideal of the 
$SU(N)$ quotient $H_{Nn}^{(N)}(\beta)$
of the Hecke algebra, the basis elements
being in one-to-one correspondence
with $SU(N)$ walk diagrams. For the sake of simplicity, we will start
with a detailed study of the $SU(3)$ meanders, before going to the
general $N$ case.

\newsec{SU(3) meander determinant}
\par
In this section, we generalize the concept of meander
to $SU(3)$ through the walk diagram picture.
A generalized meander is a couple of 
closed paths (or walk diagrams) starting and ending at the origin
of the Weyl chamber for the $sl(3)$ Lie algebra. 
The bilinear form is provided
by the standard scalar product of the Hecke algebra.
The $SU(3)$ meander determinant is obtained by an explicit
Gram-Schmidt orthogonalization of the walk-diagram basis
of a certain ideal of the $SU(3)$ quotient of the Hecke algebra.
\par
\subsec{SU(3) walk diagrams}
\par
Let us denote by
$\Lambda=(\lambda_1,\lambda_2)$ the elements of
the weight lattice $P$ of the $sl(3)$ Lie algebra 
namely the linear combinations 
$\Lambda=\lambda_1 \omega_1+\lambda_2
\omega_2$, $\lambda_1,\lambda_2 \in \IZ$, of the two fundamental
weights $\omega_1,\omega_2$, with $\omega_1^2=\omega_2^2=2/3$
and $\omega_1\cdot \omega_2=1/3$.
The Weyl chamber $P_+$
is the quotient of the weight lattice by the Weyl group,
generated by the reflections w.r.t. the walls $\lambda_1=0$ and
$\lambda_2=0$. A representative is given by
\eqn\pplus{ P_+~=~\{ (\lambda_1,\lambda_2) \in P \ \ 
{\rm such} \ {\rm that} \ \lambda_1, \lambda_2 \geq 1 \} }
\fig{The simplex $\Pi_+$. The three oriented links 
correspond respectively to $\epsilon_1$ (right), $\epsilon_2$ (up, left)
and $\epsilon_3$ (down, left). 
We have also indicated the origin $(1,1)$.}{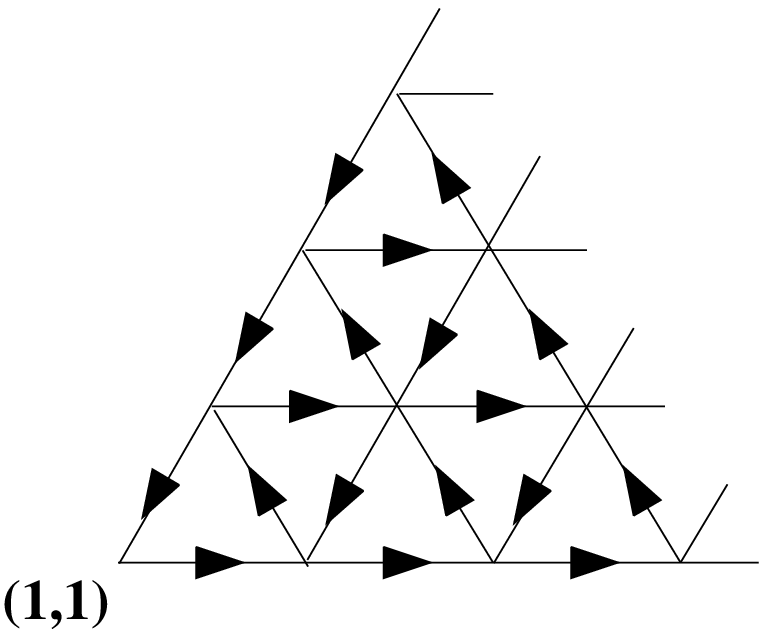}{6.cm}
\figlabel\simplex
The Weyl chamber is made into a simplex $\Pi_+$ by adding three
types of oriented bonds linking the weights (see Fig.\simplex), 
along the vectors
\eqn\vector{ \epsilon_1~=~\omega_1 \qquad \epsilon_2~=~\omega_2 -\omega_1
\qquad \epsilon_3=-\omega_2}
subject to the relation $\epsilon_1+\epsilon_2+\epsilon_3=0$.
Analogously, $P$ can be made into a simplex $\Pi$ by the same
procedure.
We define the origin of $P_+$ to be the apex $(1,1)$. 

A $SU(3)$ walk diagram of order $3n$ is an oriented closed path
of length $3n$ on $\Pi_+$, starting and ending at the origin.
It is uniquely determined by either of the following data
\item{(i)} The sequence of its $3n+1$ ``weights" in $P_+$:
$\Lambda_0=(1,1)$,
$\Lambda_1$, ..., $\Lambda_{3n-1}$, $\Lambda_{3n}=(1,1)$, such that
$\Lambda_{i+1}-\Lambda_i \in \{ \epsilon_1,\epsilon_2,\epsilon_3\}$
for $i=0,1,2,...,3n-1$. The index $i$ is referred to as the position 
of the weight $\Lambda_i$ in the sequence.
\item{(ii)} The sequence of its $3n$ ``step" vectors: 
$v_1=\epsilon_1$, $v_2$, ..., $v_{3n-1}$, $v_{3n}=\epsilon_3$
with $v_i \in \{\epsilon_1,\epsilon_2,\epsilon_3\}$ and
$(1,1)+v_1+v_2+...+v_i\in P_+$ for all $i=1,2,...,3n-1$, and
$v_1+...+v_{3n}=0$.
\par
\noindent{}The two representations are equivalent, as the
steps $v_i$ can be interpreted as $v_i=\Lambda_i-\Lambda_{i-1}$
in the sequence of weights of the walk.
\fig{A sample walk diagram of order $9$. We have indicated
by dots the successive weights visited by the path.}{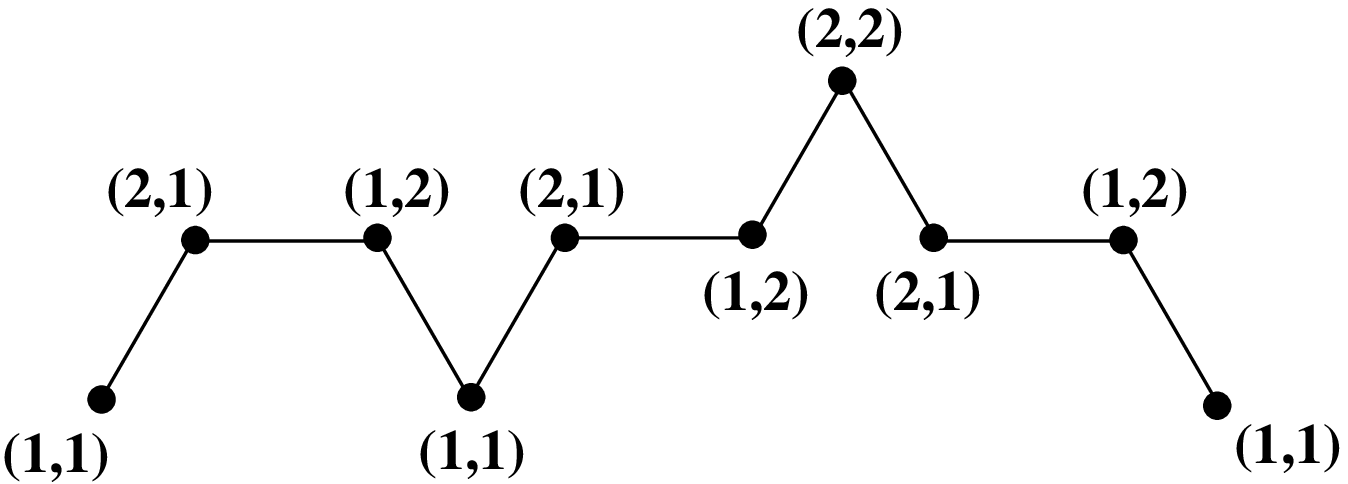}{5.cm}
\figlabel\walex

It will be useful to have a two-dimensional pictorial representation of
$SU(3)$ walk diagrams, in the same spirit as for the SU(2) walk diagrams of
the previous section (see Fig.\picto). 
We choose to represent the three possible directions
taken from each weight by three different links, with
the following correspondence:
\eqn\corsp{ \figbox{8.cm}{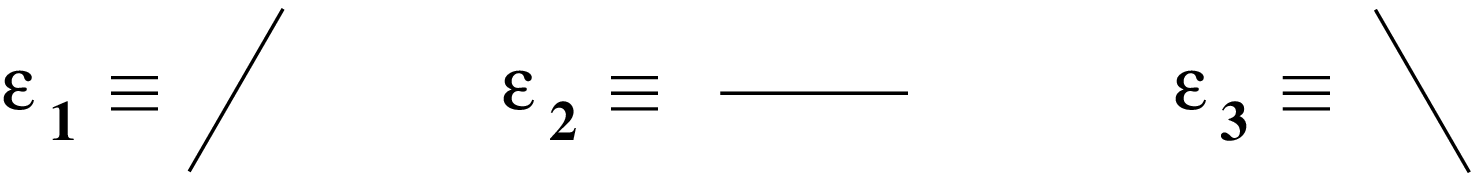} }
The walk diagrams are then represented as the corresponding succession
of these links, say from left to right 
(see the example of Fig.\walex). We denote by $W_{3n}^{(3)}$
the set of $SU(3)$ walk diagrams of order $3n$.

As a simple exercise, let us count the number $c_{3n}^{(3)}$ 
of $SU(3)$ walk diagrams of given
order $3n$. To do that, it is instructive to first count the number
$D_{(\lambda_1,\lambda_2),(\mu_1,\mu_2)}^{(N)}$ of paths
of length $N$ {\it on $\Pi$}, starting at $(\lambda_1,\lambda_2)$
and ending at $(\mu_1,\mu_2)$. As we are dealing with paths on $\Pi$,
there is no restriction other than that each step has to be
taken among $\epsilon_1,\epsilon_2,\epsilon_3$. Suppose we are taking
a total of $p$ steps $\epsilon_1$, $q$ steps $\epsilon_2$ 
and $r$ steps $\epsilon_3$, then we must have
\eqn\mutav{ \nu_1=\mu_1 -\lambda_1=p-q \ , \ 
\nu_2=\mu_2-\lambda_2=q-r \ , p+q+r=N}
hence
\eqn\valpqr{ p={N+2 \nu_1 + \nu_2 \over 3} \ , \ 
q={N-\nu_1 + \nu_2 \over 3} \ ,
r={N-\nu_1 -2 \nu_2 \over 3}}
only valid for $N+2 \nu_1+\nu_2=0$ mod 3 (there is no such path
otherwise). The number of paths is therefore equal to the number of
choices of these $p,q,r$ vectors among $N$, hence
\eqn\resuld{ D_{(\lambda_1,\lambda_2),(\mu_1,\mu_2)}^{(N)}~=~
{ N! \over  \big({N+2 \nu_1 + \nu_2 \over 3}\big)!
\big({N-\nu_1 + \nu_2 \over 3}\big)!
\big({N-\nu_1 -2 \nu_2 \over 3}\big)!} }
where $\nu_1=\mu_1-\lambda_1$, $\nu_2=\mu_2-\lambda_2$. Note that by 
translational invariance of $\Pi$
\eqn\trinv{D_{(\lambda_1,\lambda_2),(\mu_1,\mu_2)}^{(N)} 
~=~ D_{(0,0),(\nu_1,\nu_2)}^{(N)} ~\equiv~ D_{(\nu_1,\nu_2)}^{(N)}}
where we drop the origin $(0,0)$ in the last shorthand notation.
Let us now compare the paths of length $N=3n$,
from $(1,1)$ to itself, on $\Pi_+$ and on $\Pi$. On the latter simplex,
the paths can freely cross the walls of the Weyl chamber, hence there
are many more of them than on $\Pi_+$. But the latter are obtained by
reflecting any path on $\Pi$ w.r.t. the walls of the Weyl chamber,
in order to bring it back in $P_+$. Multiple reflections may be
needed to achieve this. This will eventually lead to a surjective
map from the paths on $\Pi$ to those on $\Pi_+$. 
To enumerate the $c_{3n}^{(3)}$ paths on $\Pi_+$, we have to start from those
on $\Pi$, then subtract those which cross the walls of the Weyl chamber.
Denoting by $s_1$ and $s_2$ the reflections 
w.r.t. the walls $\lambda_2=0$ and $\lambda_1=0$, we have
\eqn\reflectroi{\eqalign{ 
s_1(\lambda_1,\lambda_2)~&=~(\lambda_1+\lambda_2,-\lambda_1)\cr
s_2(\lambda_1,\lambda_2)~&=~(-\lambda_2,\lambda_1+\lambda_2)\cr
s_2s_1(\lambda_1,\lambda_2)~&=~(\lambda_1,-\lambda_1-\lambda_2)\cr
s_1s_2(\lambda_1,\lambda_2)~&=~(-\lambda_1-\lambda_2,\lambda_2)\cr
s_1s_2s_1(\lambda_1,\lambda_2)~&=~(-\lambda_2,-\lambda_1)\cr}}
which together with the identity form the six elements of the Weyl
group of $sl(3)$ (identified with the permutation group of three
objects $S_3$).
Hence the six possible reflections of the origin $(1,1)$ read
\eqn\reflor{ (1,1) \quad (-1,2) \quad (2,-1)\quad (1,-2)\quad (-2,1)
\quad (-1,-1)}
The correct subtraction formula reads
\eqn\subfor{\eqalign{ c_{3n}^{(3)}~&=~ \sum_{\sigma \in S_3}
(-1)^{l(\sigma)} D_{\sigma(1,1),(1,1)}^{(3n)}\cr
&=~D_{(0,0)}^{(3n)}
-D_{(2,-1)}^{(3n)}-D_{(-1,2)}^{(3n)}+D_{(3,0)}^{(3n)}+D_{(0,3)}^{(3n)}
-D_{(2,2)}^{(3n)}\cr
&=~ 2 {(3n)! \over (n+2)! (n+1)! n!} \cr}}
where the alternate sum ($l(\sigma)$ is the length of the permutation
$\sigma$, counting the number of reflections w.r.t. walls) accounts for
the subtraction of all the paths crossing the walls
$\lambda_2=0$ and $\lambda_1=0$, avoiding oversubtracting. 
The formula for $c_{3n}^{(3)}$ is a direct generalization of that for
the Catalan numbers \catal\ which count the number
of $SU(2)$ walk diagrams of order $2n$.  
The first few numbers $c_{3n}^{(3)}$ are listed in Table III.

$$\vbox{\font\bidon=cmr8 \def\bidondon{\bidon} \bidondon
\offinterlineskip
\halign{\tv \quad # \tv &
\hfill \ # & \hfill # & \hfill # &  \hfill #
& \hfill # &  \hfill # &  \hfill #
&  \hfill # 
\tv \cr
\noalign{\hrule}
\tvi  $n$& 1 \hfill & 2 \hfill
& 3 \hfill & 4 \hfill & 5 \hfill
& 6 \hfill & 7 \hfill & 8 \hfill
\cr
\noalign{\hrule}
\tvi $c_{3n}^{(3)}$ & 1 & 5 & 42 & 462 & 6006 & 87516 &  1385670 & 
23371634
\cr
\noalign{\hrule}
}}$$
\leg{{\bf Table III:} The numbers $c_{3n}^{(3)}$ 
of $SU(3)$ walk diagrams
of order $3n$, for $n=1,2,...,8$.}

For later use, let us also derive a formula for the numbers
$C_{(\lambda_1,\lambda_2)}^{(N)}$ of paths of $N$ steps
on $\Pi_+$, starting at the origin $(1,1)$ and ending at the weight 
$(\lambda_1,\lambda_2)\in P_+$. 
It is clear that $C_{(\lambda_1,\lambda_2)}^{(N)}$
vanishes unless $N+2 \lambda_1+\lambda_2=0$ mod $3$.
The computation is strictly analogous to that of 
$c_{3n}^{(3)}=C_{(1,1)}^{(3n)}$: we must subtract from the corresponding
paths on $\Pi$, 
$D_{(1,1),(\lambda_1,\lambda_2)}^{(N)}=D_{(\lambda_1-1,\lambda_2-1)}^{(N)}$, 
the ones which cross the walls of the Weyl chamber, resulting in 
\eqn\refty{\eqalign{ C_{(\lambda_1,\lambda_2)}^{(N)}~&=~
\sum_{\sigma \in S_3} (-1)^{l(\sigma)}
D_{\sigma(1,1),(\lambda_1,\lambda_2)}^{(N)}\cr
&=~
D_{(\lambda_1-1,\lambda_2-1)}^{(N)}-
D_{(\lambda_1+1,\lambda_2-2)}^{(N)}
-D_{(\lambda_1-2,\lambda_2+1)}^{(N)} \cr
&+D_{\lambda_1+3,\lambda_2)}^{(N)}
+D_{(\lambda_1,\lambda_2+3)}^{(N)}
-D_{(\lambda_1+1,\lambda_2+1)}^{(N)} \cr
&={\lambda_1\lambda_2(\lambda_1+\lambda_2)N! \over   
\big({N+2\lambda_1+\lambda_2\over 3}+1\big)! 
\big({N-\lambda_1+\lambda_2\over 3}+1\big)!
\big({N-\lambda_1-2\lambda_2\over 3}+1\big)!}}}

\subsec{$SU(3)$ quotient and ideal of the Hecke algebra}
\par
As mentioned above, we will now concentrate on the $SU(3)$ quotient
of the Hecke algebra, obtained by adding to the relations
\hec\ the vanishing of all Young antisymmetrizers of order $4$, 
which take the simple form
\eqn\youtroi{ Y(e_i,e_{i+1},e_{i+2})~=~Y(e_i,e_{i+1})(e_{i+2}-\mu_2)
Y(e_i,e_{i+1})~=~0 }
for $i=1,2,...,n-3$.  Noting that
\eqn\norma{ Y(e_i,e_{i+1})^2~=~(\mu_1^2 \mu_2)^{-1} Y(e_i,e_{i+1}) }
(see \idmpo),
the vanishing of \youtroi\ translates into
\eqn\minirela{ (e_i e_{i+1} e_{i+2} -e_{i+2}-e_i) Y(e_i,e_{i+1})~=~0}
for all $i=1,2,...,n-3$.

The notion of reduced element has to be slightly generalized
for $H_n^{(3)}(\beta)$ and the higher Hecke quotients.  
Indeed, the relations \hec\ and \minirela\ can be used repeatedly to reduce
any element of $H_n^{(3)}(\beta)$ to a linear combination of ``reduced
elements", which take the form of products of $e_i$'s with the
smallest possible number of factors.  However, if we try to enumerate
these reduced elements, we find non-trivial vanishing linear
combinations between them. For instance, due to \hec, we have
$e_i e_{i+1} e_i - e_i+ e_{i+1}-e_{i+1} e_i e_{i+1}=0$.
It turns out that the notion of reduced element is better 
(and usually) defined in terms
of the generators $T_i=q^{1/2}( q^{1/2} -e_i)$ mentioned above,
thanks to the relation $T_iT_{i+1} T_i=T_{i+1}T_i T_{i+1}$, as
the products of $T_i$'s with the smallest numbers of factors.
This alternative description replaces the above unwanted linear combinations 
by identities between various reduced elements, which can therefore be 
easily enumerated.  However, in view of the $SU(2)$ case, we must insist
here on working with the $e_i$'s instead of the $T_i$'s, and we
will construct a basis of $H_n^{(3)}(\beta)$ made only of reduced elements
in the $e_i$'s (this will be done in all generality in Sect.5).

Our immediate task however is not to construct a general basis 
of $H_n^{(3)}(\beta)$ but rather of a particular ideal of 
$H_{3n}^{(3)}(\beta)$. By analogy with the $SU(2)$ case, let us
consider the left
ideal ${\cal I}_{3n}^{(3)}(\beta)$ of $H_{3n}^{(3)}(\beta)$
generated by the element
\eqn\elger{Y^{(3)}_{3n}~=~Y(e_1,e_2) Y(e_4,e_5) ... Y(e_{3n-2},e_{3n-1}) }
Let us now construct a basis of reduced elements of this 
ideal using the $SU(3)$ walk
diagrams of order $3n$. 
By reduced element we mean here a product of $e_i$'s
times $Y_{3n}^{(3)}$, with the smallest number of factors.

Like in the $SU(2)$ case, let us first reexpress the walk 
diagrams of $W_{3n}^{(3)}$ in terms of box additions. We start
from the fundamental $SU(3)$ walk diagram $a_0^{(3)}$, with weights
\eqn\fundawa{\eqalign{ 
\Lambda_0~&=~(1,1)~=~\Lambda_3~=~...~=~\Lambda_{3n} \cr
\Lambda_1~&=~(2,1)~=~\Lambda_4~=~...~=~\Lambda_{3n-2} \cr
\Lambda_2~&=~(1,2)~=~\Lambda_5~=~...~=~\Lambda_{3n-1} \cr}}
This is the most compact path of length $3n$ on $\Pi_+$.
In the abovementioned pictorial representation $a_0^{(3)}$
reads
\eqn\picano{ a_0^{(3)}~=~\figbox{5.cm}{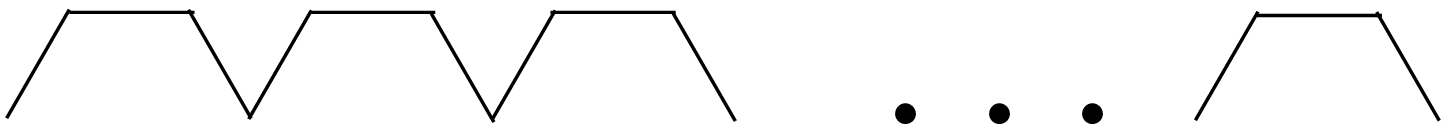} }
For a general walk $a\in W_{3n}^{(3)}$, we define three types of box
additions at position $i$, still denoted by $a\to a+\diamond_i$,
according to the configuration of the weights $\Lambda_{i-1}$,
$\Lambda_i$, $\Lambda_{i+1}$ of $a$ at positions
$i-1$, $i$ and $i+1$:
\item{(i)} $\Lambda_{i+1}-\Lambda_i=\epsilon_1$ and 
$\Lambda_i-\Lambda_{i-1}=\epsilon_2$. A box addition at position $i$
transforms $\Lambda_i \to \Lambda_i+\epsilon_1-\epsilon_2$ and leaves
all the other weights unchanged.
\item{(ii)} $\Lambda_{i+1}-\Lambda_i=\epsilon_1$ and 
$\Lambda_i-\Lambda_{i-1}=\epsilon_3$. A box addition at position $i$
transforms $\Lambda_i \to \Lambda_i+\epsilon_1-\epsilon_3$ and leaves
all the other weights unchanged.
\item{(iii)} $\Lambda_{i+1}-\Lambda_i=\epsilon_2$ and
$\Lambda_i-\Lambda_{i-1}=\epsilon_3$. A box addition at position $i$
transforms $\Lambda_i \to \Lambda_i+\epsilon_2-\epsilon_3$ and leaves
all the other weights unchanged.

\noindent{}Pictorially, this is summarized by the following box additions,
according to the case at hand. 
\eqn\posicas{ (i):\ \figbox{1.2cm}{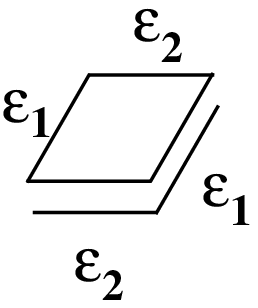}\ \ \ (ii):\ \figbox{.8cm}{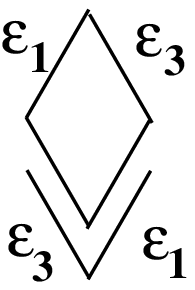}
\ \ \ (iii):\ \figbox{1.2cm}{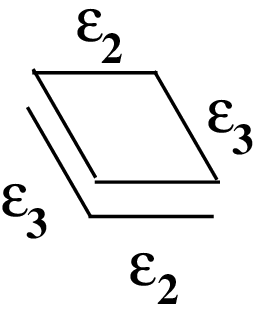} }
If the weights of $a$ are not in one of the three cases $(i)-(iii)$ above,
the box addition cannot be performed at the position $i$.
For instance, on the fundamental walk $a_{0}^{(3)}$, box additions
can be performed only at positions $3,6,9,...,3n-3$, and fall in 
the case $(ii)$.
This construction gives a procedure to describe any $SU(3)$ walk
diagram as a sequence of box additions on the fundamental
walk $a_0^{(3)}$. This description is however not unique, as different
sequences may lead to the same walk diagram.
The order in which the box additions are made is not a problem, the
only difficulty here is the occurence of hexagons in the box decomposition
of $a$ (i.e. the filling of the space between $a_0^{(3)}$ and $a$ with 
boxes of type $(i)-(iii)$), because there are two different ways of
filling an hexagon with boxes, namely
\eqn\fibohe{ \figbox{1.2cm}{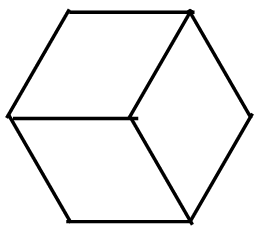} \qquad  {\rm or} \qquad  
\figbox{1.2cm}{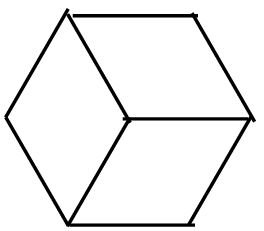}}
To fix this ambiguity, we simply forbide any box addition on $a$ 
which would create an hexagon of the second type in \fibohe, namely
we do not allow the following box addition at position $i$
\eqn\forbid{ \figbox{1.2cm}{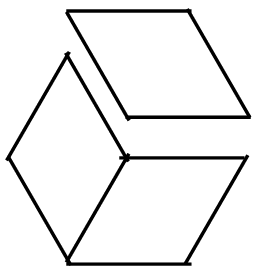} }
With this latter rule, each walk diagram $a\in W_{3n}^{(3)}$
has a unique box 
decomposition,
namely a non-ordered sequence of box additions to be performed on 
$a_0^{(3)}$ leading to $a$. Such a box decomposition can be 
pictorially represented by filling the space between $a_0^{(3)}$
and $a$ with the corresponding boxes. 

We are now ready to establish a map $\varphi$ between $W_{3n}^{(3)}$ and
the reduced elements of ${\cal I}_{3n}^{(3)}(\beta)$.
We start with the fundamental walk
\eqn\mapfun{ \varphi(a_0^{(3)})~=~Y_{3n}^{(3)} }
defined in \elger, and construct all the other reduced elements
by induction on box additions, namely
\eqn\reboa{ \varphi(a+\diamond_i)~=~ e_i \varphi(a)}
for all $a\in W_{3n}^{(3)}$. 
This expression is well-defined, as at each step, the various box
additions which can be performed say at positions $i$ and $j$
on a diagram $a$ satisfy 
$|j-i|>1$, hence the corresponding $e_i$ and $e_j$
commute: the order of their left multiplication does not matter. 
Moreover, we have taken care of the hexagon ambiguities\foot{
Having forbidden all the hexagons \forbid, the only hexagons
appearing in the box decomposition of any $a\in W_{3n}^{(3)}$
are of the form of the first hexagon of \fibohe. We could have 
decided to make a more symmetric choice for $\varphi$, namely
by associating the combination $Y(e_i,e_{i+1})$
instead of $e_ie_{i+1}e_i$ to each of these hexagons (which would
then be represented empty, without their inner box decomposition). 
This however
would not affect the final value of the meander determinant, 
allowing us to stick to our non-symmetric choice.
The symmetric choice would have the only advantage of putting
the hexagons in the box decomposition on the same footing as
those over which the walk rests (i.e., forming the product
$Y_{3n}^{(3)}$, defining the ideal).} 
by forbiding \forbid.
This leads to the definition of the basis 1 of ${\cal I}_{3n}^{(3)}(\beta)$,
with elements
\eqn\basonet{ (a)_1~=~ (\mu_1^2 \mu_2)^{n\over 2}\ \varphi(a)\ , \ 
a\in W_{3n}^{(3)} }
(The choice of the normalization factor will become clear below.).
As an immediate consequence the vector space ${\cal I}_{3n}^{(3)}(\beta)$
has dimension\foot{Let us stress that this basis is distinct from 
the standard basis of \DJ\ \MUR, when restricted to the
ideal ${\cal I}_{3n}^{(3)}(\beta)$. The latter uses indeed the generators
$T_i$ \genet. Our non-standard choice finds its justification
in the $SU(2)$ case, in which meanders are recovered.} 
$c_{3n}^{(3)}$ \subfor.

Let us illustrate this construction with the case $n=2$.
There are $c_{6}^{(3)}=5$ walk diagrams, and the basis 1 of
${\cal I}_{6}^{(3)}(\beta)$ reads
\eqn\bsoi{\eqalign{
\left( \figbox{2.cm}{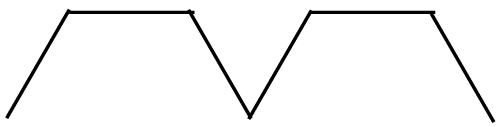} \right)_1~&=~
\mu_1^2 \mu_2   Y(e_1,e_2) Y(e_4,e_5) \cr
\left( \figbox{2.cm}{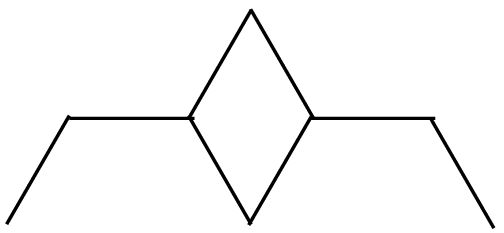} \right)_1~&=~
\mu_1^2 \mu_2   e_3 Y(e_1,e_2) Y(e_4,e_5) \cr
\left( \figbox{2.cm}{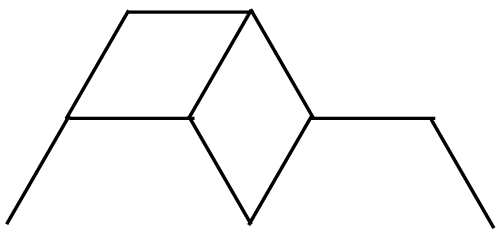} \right)_1~&=~
\mu_1^2 \mu_2   e_2 e_3 Y(e_1,e_2) Y(e_4,e_5) \cr
\left( \figbox{2.cm}{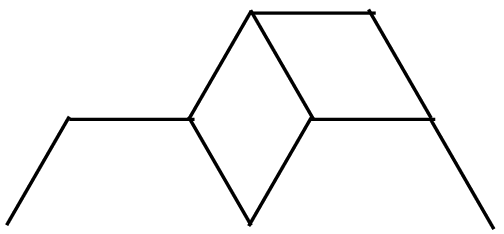} \right)_1~&=~
\mu_1^2 \mu_2   e_4 e_3 Y(e_1,e_2) Y(e_4,e_5) \cr
\left( \figbox{2.cm}{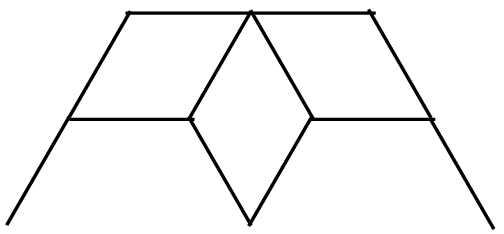} \right)_1~&=~
\mu_1^2 \mu_2   e_2 e_4 e_3 Y(e_1,e_2) Y(e_4,e_5) \cr }}
where we have represented, for each walk diagram, 
the box additions performed on the fundamental one (the box decompositions).
It is instructive to recover the basis \bsoi\ by a direct study of
the left ideal 
${\cal I}_{6}^{(3)}(\beta)=H_{6}^{(3)}(\beta) Y_6^{(3)}$, with 
$Y_6^{(3)}$ as in \elger.
Noting that 
\eqn\propgt{ e_i Y(e_i,e_{i+1})~=~e_{i+1} Y(e_i,e_{i+1})~=~\beta
Y(e_i,e_{i+1}) }
we see that the only new element of ${\cal I}_{6}^{(3)}(\beta)$
obtained by acting with one $e_i$ on the fundamental one $Y_6^{(3)}$
is $e_3 Y_6^{(3)}$ confirming the fact that the only possible
box addition on $a_0^{(3)}$ here is at position $i=3$. 
Acting with two extra $e_i$'s, we easily find the only other elements
$e_2 e_3 Y_6^{(3)}$, $e_4 e_3 Y_6^{(3)}$ and $e_2 e_4 e_3 Y_6^{(3)}$.
This exhausts all reduced elements of ${\cal I}_6^{(3)}(\beta)$, as
$e_3 e_2 e_4 e_3 Y_6^{(3)}=(\beta e_3+(e_2+e_4)e_3-2)Y_6^{(3)}$.
Note that we still have not met here any hexagon ambiguity,
occurring only for $n\geq 3$.
For completeness,
we list below the box decompositions relevant to the $n=3$ case,
with $c_{9}^{(3)}=42$ walk diagrams:
\eqn\lisbi{\figbox{9.cm}{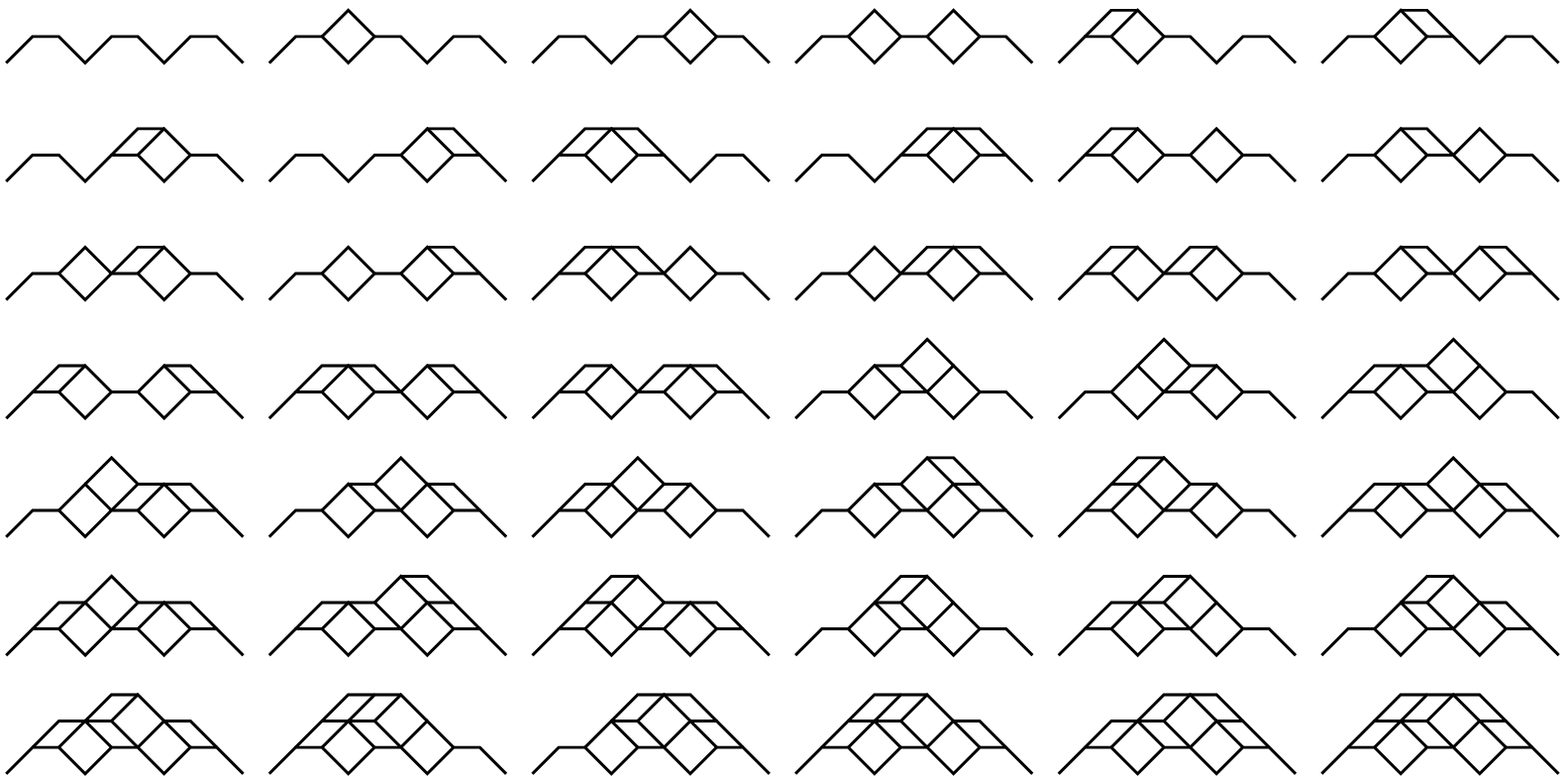} }
Note that the third diagram in the first line of \lisbi\ 
is nothing but the box decomposition of the sample
walk of Fig.\walex. Note also that only hexagons of the first type
of \fibohe\ appear in the above box decompositions.

A $SU(3)$ meander is now identified as a pair $(a)_1,(b)_1$
of basis 1 elements for 
${\cal I}_{3n}^{(3)}(\beta)$. Following the $SU(2)$ example,
we may attach to each meander the quantity 
$\big( (a)_1,(b)_1\big)={\rm Tr}\big((a)_1(b)_1^t\big)$.
Here, the Markov trace on $H_{3n}^{(3)}$ (still denoted by ${\rm Tr}$)
is normalized so that 
\eqn\normatroi{{\rm Tr}(1)~=~U_2(\beta)^{3n}~=~(\beta^2-1)^{3n}} 
and still defined by induction through the relation \marko, but with a 
different constant $\eta$, namely
\eqn\etacho{ \eta~=~ \mu_2~=~ {\beta \over \beta^2-1}}
Let us consider the Gram matrix of the basis 1, with
entries
\eqn\gramatr{ \big[{\cal G}_{3n}^{(3)}(\beta)\big]_{a,b}~=~
\big( (a)_1,(b)_1\big)}
As an example, the Gram matrix for $n=2$ reads (with the same ordering
of the basis elements as in \bsoi)
\eqn\ntwoex{{\cal G}_6^{(3)}(\beta)~=~\b^2(\b^2-1)\pmatrix{
\b^2-1 & \b   & \b^2   & \b^2   & \b^3 \cr
\b   & \b^2   & 2 \b   & 2 \b   & \b^2+2\cr
\b^2 & 2 \b   & 2 \b^2 & \b^2+2 & \b(\b^2+2)\cr
\b^2 & 2 \b   & \b^2+2 & 2 \b^2 & \b(\b^2+2)\cr
\b^3 & \b^2+2 & \b(\b^2+2) & \b(\b^2+2) & \b^2(\b^2+2)\cr}}

\subsec{$SU(3)$ meander determinant: main result}
\par 
We define the $SU(3)$ meander determiant 
$\Delta_{3n}^{(3)}(\beta)$ as the determinant of the Gram matrix
\gramatr\ of the
basis 1 of ${\cal I}_{3n}^{(3)}(\beta)$. 
The aim of this section is to prove the following formula for 
this determinant
\eqn\fordetroi{ \Delta_{3n}^{(3)}(\beta)~=~\prod_{m=1}^{n+1}
\big[U_m(\beta)\big]^{a_{m,n}^{(3)}} }
where
\eqn\aleq{\eqalign{ a_{m,n}^{(3)}~&=~
\sum_{\sigma\in S_3} (-1)^{l(\sigma)} C_{(m+1,m+1)-\sigma(1,1)}^{(3n)}\cr
&=~C_{(m,m)}^{(3n)}
-C_{(m+2,m-1)}^{(3n)}-C_{(m-1,m+2)}^{(3n)} \cr
&+C_{(m+3,m)}^{(3n)}+C_{(m,m+3)}^{(3n)}
-C_{(m+2,m+2)}^{(3n)} \cr}}
for $m\geq 2$ and
\eqn\seqalp{ a_{1,n}^{(3)}~=~C_{(4,1)}^{(3n)}+C_{(1,4)}^{(3n)}
-C_{(3,3)}^{(3n)} }
in terms of the numbers $C_{(\lambda_1,\lambda_2)}^{(N)}$ of paths
of length $N$ on $\Pi_+$ starting from the origin $(1,1)$ and terminating
at $(\lambda_1,\lambda_2)$, computed in \refty\ (it is understood that
$C_{(\lambda_1,\lambda_2)}^{(3n)}$ vanishes  
unless $(\lambda_1,\lambda_2)\in P_+$).
The first few values of the powers $a_{m,n}^{(3)}$ of
$U_m$ in the $SU(3)$ meander determinant $\Delta_{3n}^{(3)}(\beta)$
are listed in Table IV.

$$\vbox{\font\bidon=cmr8 \def\bidondon{\bidon} \bidondon
\offinterlineskip
\halign{\tv \quad # \tv &
\hfill \ # & \hfill # & \hfill # &  \hfill #
& \hfill # &  \hfill # &  \hfill #  & \hfill #
\tv \cr
\noalign{\hrule}
\tvi  $\scriptstyle m\backslash n$& 1 \hfill & 2 \hfill
& 3 \hfill & 4 \hfill & 5 \hfill
& 6 \hfill & 7 \hfill & 8 \hfill\cr
\noalign{\hrule}
\tvi 1 & 1 & 6 & 42 & 297 & 1430  & -14586 & -764218 & -21246940\cr
\tvi 2 & 1 & 6 & 63 & 814 & 11583 & 175032 & 2762942 & 45108888\cr
\tvi 3 &   & 4 & 42 & 506 & 7306  & 119340 & 2098208 & 38571368\cr
\tvi 4 &   &   & 21 & 374 & 5707  & 89352 & 1495490  & 26803832\cr
\tvi 5 &   &   &   & 121  & 3276  & 65790 & 1218356  & 22309287\cr
\tvi 6 &   &   &   &   &    728   & 27336 & 701879   & 15622750\cr
\tvi 7 &   &   &   &   &          & 4488  & 218994   & 6931694\cr
\tvi 8 &   &   &   &   &        &         & 28101    & 1701678\cr
\tvi 9 &   &   &   &   &        &         &          & 177859\cr
\noalign{\hrule}
}}$$
\leg{{\bf Table IV:} The powers $a_{m,n}^{(3)}$ of $U_m$ in
the $SU(3)$ meander determinant of order $3n$,
$\Delta_{3n}^{(3)}(\beta)$, for $n=1,2,...,8$.}

The formula \fordetroi\ exhibits a remarkable feature: the numbers 
$a_{m,n}^{(3)}$ are obtained from the $C$'s by the same 
addition/subtraction formula as that giving the $C$'s in 
terms of the $D$'s, namely between the numbers of paths 
on $\Pi_+$ and those on $\Pi$ (see \refty). This feature was already
present in the $SU(2)$ case, if we note that 
eqs.\detforme\-\pathnum\ translate into
\eqn\transtwo{\eqalign{ 
C_{2m+1}^{(2n)}~&=~D_{2m}^{(2n)} -
D_{2m+2}^{(2n)} \cr
a_{m,n}^{(2)}~&=C_{2m+1}^{(2n)}-C_{2m+3}^{(2n)} \cr}}
where 
we have introduced the numbers $D_{2m}^{(2n)}={2n \choose
n-m}$ of paths of $2n$ steps 
from $0$ to $2m$ on the integer line $\IZ$, identified with
the weight lattice $P$ of $SU(2)$.

The validity of \fordetroi\ is readily checked in the case $n=2$, 
where we find (see also Table IV), 
by direct computation of the determinant of \ntwoex\
\eqn\detntoex{ \Delta_6^{(3)}(\beta)~=~U_1^6 U_2^6 U_3^4}
in agreement with $C_{(1,1)}^{(6)}=C_{(5,2)}^{(6)}=C_{(1,4)}^{(6)}=5$, 
$C_{(4,1)}^{(6)}=10$,
$C_{(2,2)}^{(6)}=16$, $C_{(2,5)}^{(6)}=0$ and $C_{(3,3)}^{(6)}=9$. 

The formula \fordetroi\  is proved below, by the explicit
Gram-Schmidt orthogonalization of the basis 1, namely
the construction of a new basis (which we call basis 2), still
indexed by the walk diagrams $a\in W_{3n}^{(3)}$, and such that
\eqn\eletwo{ (a)_2~=~\sum_{b\in W_{3n}^{(3)}\atop
b\subset a} P_{a,b} (b)_1 }
where the sum extends over the walk diagrams $b$ ``included" in $a$,
namely such that $a$ can be obtained from $b$ by box additions.
The basis 2 is orthonormal w.r.t. the scalar product $(\ , \ )$,
namely $\big( (a)_2,(b)_2\big)=\delta_{a,b}$ for any $a,b\in W_{3n}^{(3)}$.

\subsec{The orthonormal basis}

The orthonormal basis 2 is constructed as follows.
We start with the fundamental element
\eqn\fonto{\eqalign{ (a_0^{(3)})_2~&=~(\mu_1^2 \mu_2)^{n/2} (a_0^{(3)})_1~
=~(\mu_1^2\mu_2)^{n} Y_{3n}^{(3)}\cr 
&=~y(e_1,e_2)y(e_4,e_5)...y(e_{3n-2},e_{3n-1})\cr}}
with $Y_{3n}^{(3)}$ as in \elger. The last line of \fonto\
is a reexpression in terms of 
the idempotent antisymmetrizers of order $3$ of 
\propY\-\procons.
The normalization factor in \fonto\
ensures that $(a_0^{(3)})_2$ has norm $1$, as
$\big( (a_0^{(3)})_1,(a_0^{(3)})_1\big)
={\rm Tr}(Y_{3n}^{(3)})=(\mu_1^2 \mu_2)^{-n}$,
by immediate application of the Markov property \marko.
The other basis 2 elements are constructed by box additions 
on the fundamental one,
through the recursion relation 
\eqn\boxredf{ (a+\diamond_{i,m})_2~=~\sqrt{\mu_{m+1}\over \mu_m}
(e_i-\mu_m)\ (a)_2}
where $m$ is the ``height" of the box, defined as
\eqn\defheig{ m~=~\Lambda_i \cdot (\Lambda_{i+1}+\Lambda_{i-1}-2 \Lambda_i)}
in terms of the weights $\Lambda_{i-1}$, $\Lambda_i$,
$\Lambda_{i+1}$ of $a$ with respective positions $i-1,i,i+1$, for all
$1 \leq i\leq 3n-1$ (note that the effect of the box addition
at position $i$ is to change 
$\Lambda_i\to \Lambda_{i+1}+\Lambda_{i-1}-\Lambda_i$ 
in all cases (i)-(iii) of \posicas).
Note that
as \boxredf\ depends explicitly on the height $m$ of the box addition,
we have added the subscript $m$ to the box symbol $\diamond_{i,m}$.
For simplicity, we will denote by $(\diamond_{i,m})$ the element of the 
Hecke algebra which multiplies $(a)_2$ in \boxredf.
The fundamental property of \boxredf\ is that it resolves the
hexagon ambiguity, namely the two ways \fibohe\ of building an hexagon
by these box additions are equivalent, i.e.\foot{This equation takes
the form of the celebrated Yang-Baxter equation for the so-called
trigonometric limit of the $A_2$ RSOS model of \JM.},
\eqn\twoeq{  (e_i -\mu_m)(e_{i+1}-\mu_{m+p+1})(e_i-\mu_p)~=~
(e_{i+1}-\mu_p)(e_i-\mu_{m+p+1})(e_{i+1}-\mu_m)}
where $m=\Lambda\cdot (\epsilon_1-\epsilon_2)-1$ and
$p=\Lambda\cdot (\epsilon_2-\epsilon_3)-1$ and $\Lambda$ denotes the
weight of the leftmost vertex of the hexagons. Eq.\twoeq\ is easily proved
by using the definition \nota\ for $\mu$ and the recursion relation
for the Chebishev polynomials 
$U_{m+1}=\beta U_m-U_{m-1}$, together with the Hecke algebra relations
\hec.
We also note that, upon defining $\mu_0=0$,
the box additions \boxredf\ enable us to rewrite
each of the building blocks $Y(e_i,e_{i+1})$ of the fundamental element 
$(a_0^{(3)})_2$ as the hexagonal result of three box additions on
an empty walk diagram. For each such  
hexagon, the equivalence \twoeq\ simply amounts
to $Y(e_i,e_{i+1})=Y(e_{i+1},e_i)$. 
In this way, any basis 2 element can be seen
as the result of box additions on the vacuum diagram (identified 
with the unit 1 of $H_{3n}^{(3)}(\beta)$) as well.

In the example $n=2$ of \bsoi, we have the basis 2 elements
\eqn\bsot{\eqalign{
\left( \figbox{2.cm}{ba1.eps} \right)_2~&=~
\mu_1^4 \mu_2^2 Y(e_1,e_2) Y(e_4,e_5) \cr
\left( \figbox{2.cm}{ba2.eps} \right)_2~&=~
\mu_1^4 \mu_2^{3/2} \mu_3^{1/2} (e_3-\mu_2) Y(e_1,e_2) Y(e_4,e_5) \cr
\left( \figbox{2.cm}{ba3.eps} \right)_2~&=~
\mu_1^{7/2} \mu_2^2 \mu_3^{1/2} 
(e_2-\mu_1) (e_3-\mu_2) Y(e_1,e_2) Y(e_4,e_5) \cr
\left( \figbox{2.cm}{ba4.eps} \right)_2~&=~
\mu_1^{7/2} \mu_2^2\mu_3^{1/2} 
(e_4-\mu_1) (e_3-\mu_2) Y(e_1,e_2) Y(e_4,e_5) \cr
\left( \figbox{2.cm}{ba5.eps} \right)_2~&=~
\mu_1^3 \mu_2^{5/2} \mu_3^{1/2} (e_2-\mu_1) (e_4-\mu_1) 
(e_3-\mu_2) Y(e_1,e_2) Y(e_4,e_5) \cr }}

The basis 2, determined by \fonto\-\boxredf, coincides 
up to normalization
factors with the so-called semi-normal basis $(a)_{sn}$ 
(still indexed by the walk diagrams $a\in W_{3n}^{(3)}$)
of \DJ\ \MUR\ when restricted to the ideal ${\cal I}_{3n}^{(3)}(\beta)$
(identified with the top Specht module of $H_{3n}^{(3)}(\beta)$). 
The semi-normal basis elements $(a)_{sn}$ for ${\cal I}_{3n}^{(3)}$
satisfy stronger relations, 
namely that
\eqn\strong{ (a)_{sn}^t (b)_{sn}~=~0 \quad {\rm unless}\ a~=~b}
Let us assume the analogous relations for the basis 2 elements, and
verify that all the $(a)_2$ have norm $1$. Let us rewrite
the quantity $(a+\diamond_i)_2^t (a+\diamond_i)_2$, by ``moving the box"
from the left factor to the right: this amounts to an extra left
multiplication by $(e_i-\mu_m)^t=(e_i-\mu_m)$ on $(a)_2$, namely
\eqn\simuop{\eqalign{ (a+\diamond_i)_2^t (a+\diamond_i)_2~&=~
(a)_2^t {\mu_{m+1}\over \mu_m}(e_i-\mu_m)^2 (a)_2\cr
&=~(a)_2^t {\mu_{m+1}\over \mu_m} \big( (\mu_1^{-1}-2\mu_m)(e_i-\mu_m)+
\mu_m(\mu_1^{-1}-\mu_m) \big) (a)_2\cr
&=~({1 \over \sqrt{\mu_m\mu_{m+1}}}-\sqrt{\mu_m\mu_{m+1}})(a)_2^t 
(a+\diamond_i)_2   +(a)_2^t (a)_2 \cr
&=~ (a)_2^t (a)_2\cr}}
where we have first used $e_i^2=\beta e_i=\mu_1^{-1}e_i$, then
the recursion relation $\mu_1^{-1}-\mu_m=\mu_{m+1}^{-1}$.
We have dropped the term proportional to 
$(a)_2^t(a+\diamond_i)_2=0$ by application of \strong.
Now eq.\simuop, enables us to prove by induction on the box additions
that $\big((a)_2,(a)_2\big)=1$ for all $a\in W_{3n}^{(3)}$, as
$\big((a_0^{(3)})_2,(a_0^{(3)})_2\big)=1$.
This fixes the prefactor in \boxredf.

The relation \strong\ for basis 2 elements, namely 
\eqn\strop{ (a)_2^t(b)_2~=~0\qquad {\rm unless}\ \  a~=~b}
can be directly proved by induction on the number
of boxes, denoted by $|a|$ and $|b|$ 
in the box decompositions of $a$ and $b$,
in the same spirit as for the $SU(2)$ case (see \MOI).
Let us give a brief description of this proof for completeness.
The aim is to prove by induction on the integer $k$ the following property
\eqn\hypot{ {\cal P}_k\ :\ \  (a)_2^t (b)_2~=~0\ \ \ {\rm for}\ {\rm any}
\ a\neq b\in W_{3n}^{(3)}, \ {\rm with}\ |a|=k\leq |b|}
Assume that ${\cal P}_{k-1}$ is true for some $k\geq 1$. 
Let us consider a pair $a$, $b$ of walk diagrams with $|a|=k$
and $|b|\geq k$, $b\neq a$.
We write the
walk $a$ with $k$ boxes as the result of a box addition on some $a'$,
at position $i$, with height $\ell$,
namely $a=a'+\diamond_{i,\ell}$, and $|a'|=|a|-1=k-1$.
We then rewrite the product
$(a)_2^t(b)_2=(a'+\diamond_{i,\ell})_2^t (b)_2$ 
by letting this box act on $(b)_2$ by left multiplication. Three situations
may occur, according to the configuration of the weights 
$\Lambda_{i-1}$, $\Lambda_i$, $\Lambda_{i+1}$
of $b$ at respective positions 
$i-1$, $i$, $i+1$. 
Setting $\Lambda_i-\Lambda_{i-1}=\epsilon_r$ and 
$\Lambda_{i+1}-\Lambda_i=\epsilon_s$, we have the three possibilities
\item{(i)} $b$ has a minimum at 
position $i$, namely $(r,s)=(2,1)$, $(3,2)$ or $(3,1)$.
Let $m=\Lambda_i\cdot(\epsilon_s -\epsilon_r)$ be the height of the box
to be added on $b$ at position $i$, we simply rewrite
\eqn\boxrew{ (\diamond_{i,\ell})~=~\sqrt{\mu_{\ell+1}\over \mu_{\ell}}
(e_i-\mu_\ell)~=~\sqrt{\mu_{\ell+1}\mu_m\over \mu_\ell \mu_{m+1}}
(\diamond_{i,m}) +\sqrt{\mu_{\ell+1}\over \mu_\ell}(\mu_m-\mu_\ell)}
hence $(a)_2^t (b)_2$ can be reexpressed as a linear combination of
$(a')_2^t(b+\diamond_{i,m})_2$ and $(a')_2^t (b)_2$.
\item{(ii)} $b$ has a maximum at position $i$, namely $(r,s)=(1,2)$, 
$(2,3)$ or $(1,3)$. Using the equivalence \twoeq, we can always 
arrange for $(b)_2$ to be the result $(b'+\diamond_{i,m})_2$ of 
a box addition on some $b'$, with $|b'|=|b|-1$ (we also write
$(b')_2=(b-\diamond_{i,m})_2$).
Hence the element $(b')_2$ is multiplied from the left by
\eqn\mutib{ (\diamond_{i,\ell})(\diamond_{i,m})~=~
\sqrt{\mu_{\ell+1}\over \mu_\ell}
(\mu_{\ell+1}^{-1}-\mu_m)(\diamond_{i,m})+ \sqrt{\mu_{\ell+1}\mu_m\over
\mu_\ell\mu_{m+1}} }
and $(a)_2^t(b)_2$ is expressed as a linear combination of $(a')_2^t(b)_2$
and $(a')_2^t (b')_2$, with $|b'|=|b|-1$.
\item{(iii)} $b$ has a slope at position $i$, namely $r=s=1$, $2$ or $3$.
Without loss of generality, we may assume that $b$ contains the two
boxes $f=(\diamond_{i-1,m+1})(\diamond_{i,m})$. The left multiplication
by $(\diamond_{i,\ell})$ consists of two terms, one proportional
to $e_i f$, and the other proportional to $f$. The former is proportional
to
\eqn\gifo{\eqalign{ e_i(e_{i-1}-\mu_m)(e_i-\mu_{m-1})~&=~Y(e_{i-1},e_i) 
-\mu_{m-1} e_i e_{i-1} \cr
&=~\big( e_{i-1}(e_i-\mu_1) -\mu_{m-1}e_i \big) e_{i-1} \cr}}
hence $e_i$ has commuted through the two boxes, creating a left
factor of $e_{i-1}$. Now we can repeat this process, until we meet the
bottom of the diagram $b$. Several cases have to be inspected, let 
us simply give one example: we are left, say, with the left multiplication 
of $e_j$ with the bottom boxes 
$(\diamond_{j-1,1})(\diamond_{j-2,2})Y(e_{j-1},e_{j})$, which gives
\eqn\diamfo{\eqalign{ 
e_j (e_{j-1}&-\mu_1)(e_{j-2}-\mu_2) Y(e_{j-1},e_j)\cr
&=~\mu_1 Y(e_{j-1},e_j) (e_{j-2}-\mu_2) Y(e_{j-1},e_j)\cr
&=~0\cr}}
where we used $Y(e_{j-1},e_j)=\mu_1 e_j Y(e_{j-1},e_j)$, commuted 
$e_j$ through $e_{j-2}$, and finally used 
the vanishing condition of the fourth order 
antisymmetrizer \youtroi\ in $H_{3n}^{(3)}(\beta)$.
\par
\noindent{}
To summarize, in all cases (i)-(iii) above, we have been able to rewrite
$(a)_2^t(b)_2$ as a linear combination of terms of the form
$(a')_2^t (b'')_2$, where $b''=b+\diamond_i$, $b$ or $b-\diamond_i$.
In all cases, we have $|a'|=|a|-1=k-1$, and $|b''|\geq |b|-1\geq k-1$.
Moreover, $b''\neq a'$: otherwise, one would have been necessarily in the
case (ii) with $b''=b-\diamond_i=a'=a-\diamond_i$, hence $a=b$, which 
contradicts the hypothesis. Hence $b''\neq a'$ and we may apply to each
pair $(a',b'')$
the induction hypothesis ${\cal P}_{k-1}$, hence $(a')_2^t(b'')_2=0$
in all cases at hand, and ${\cal P}_k$ follows.
There remains to prove ${\cal P}_0$. We have $a=a_0^{(3)}$, the only walk
with 0 boxes. We simply have to act on the left
of $(b)_2$ with the hexagons $Y(e_i,e_{i+1})$ forming $(a)_2^t=(a)_2$.
The result vanishes for all $b\neq a_0^{(3)}$, as at least one
of these hexagons, say $Y(e_j,e_{j+1})$ has a right factor
$e_j$ or $e_{j+1}$, whose position corresponds to a slope of $b$ 
(the result of the left multiplication of $(b)_2$ by this yields zero,
like in the case (iii) above), or yields directly zero by the vanishing
of the antisymmetrizers of order 4.  
This completes the proof of ${\cal P}_k$
for all $k\geq 0$, and \strop\ follows.

As a final remark, the property \eletwo\ follows directly from the
recursive definition \boxredf. Indeed, the process of box addition
only involves walk diagrams included in $a$ for the construction of
$(a)_2$, hence the change of basis 1 $\to $ 2 is triangular.

\subsec{$SU(3)$ meander determinant: the proof}

Using \eletwo, we can easily reexpress the $SU(3)$ meander determinant
as
\eqn\medex{ \Delta_{3n}^{(3)}(\beta)~=~ \prod_{a \in W_{3n}^{(3)}}
P_{a,a}^{-2} }
as the basis 2 is orthonormal, and the change of basis 1 $\to$ 2 is
triangular, with normalization factors $P_{a,a}$ on the diagonal.
The quantities $P_{a,a}^{2}$ are easily computed by induction.
First we have, from \fonto\
\eqn\firnor{ P_{a_0^{(3)},a_0^{(3)}}^{2}~=~(\mu_1^2 \mu_2)^n }
and from \boxredf\
\eqn\recpaa{ P_{a+\diamond_{i,m},a+\diamond_{i,m}}^2~=~
{\mu_{m+1} \over \mu_m}\ P_{a,a}^2 }
for all $a\in W_{3n}^{(3)}$.
Each term $P_{a,a}^2$ is therefore expressed as a product over all the
box additions leading from $a_0^{(3)}$ to $a$
\eqn\expaa{ P_{a,a}^2~=~ (\mu_1^2 \mu_2)^n 
\prod_{{\rm box}\ {\rm additions} \atop
{\rm from}\ a_0^{(3)} \ {\rm to}\ a} {\mu_{m+1} \over \mu_m} }
where $m$ stands for the height of the box addition.

In the $n=2$ example of \bsoi\-\bsot, the five walks have respective
values of $P_{a,a}^2$
\eqn\fivwa{ \mu_1^4 \mu_2^2 \ , \ \mu_1^4 \mu_2 \mu_3 \ , \ 
\mu_1^3 \mu_2^2 \mu_3 \ , \ \mu_1^3 \mu_2^2 \mu_3 \ , \ 
\mu_1^2 \mu_2^4 \mu_3 }
leading immediately to \detntoex, using \medex.  

The product in \expaa\ can be further simplified, by noting
that the powers of $\mu$ can be redistributed to each of the
individual steps $v_i=\Lambda_{i}-\Lambda_{i-1}$ forming $a$.  
For each such step, say from $\Lambda=(\lambda_1,\lambda_2)$ to
$\Lambda'=(\lambda_1',\lambda_2')$, let us define a weight function
\eqn\weight{ w(\Lambda,\Lambda')~=~ \left\{ \matrix{
\sqrt{ \mu_{\lambda_1}\mu_{\lambda_1+\lambda_2}}
&{\rm if}\ \Lambda'-\Lambda=\epsilon_1 \cr
\sqrt{ \mu_{\lambda_2}\mu_{\lambda_1'}}
&{\rm if}\ \Lambda'-\Lambda=\epsilon_2 \cr
\sqrt{ \mu_{\lambda_2'}\mu_{\lambda_1'+\lambda_2'}}
&{\rm if}\ \Lambda'-\Lambda=\epsilon_3 \cr} \right. }
Now, by inspection of the three possible box additions \posicas, 
we see that the weights exactly follow the rule \boxredf, namely
that
\eqn\detcal{\prod_{a \in W_{3n}^{(3)}} P_{a,a}^2~=~ 
\prod_{{\rm steps}\ v \ {\rm in}\ {\rm all}
\atop {\rm walks} \ a\in W_{3n}^{(3)} } w(v) }

This enables us to identify the total power $\alpha_{m,n}^{(3)}$
of $\mu_m$ in the product
\detcal. Indeed a factor $\mu_m^{1/2}$ will appear whenever 
in a step with value $\epsilon_1$, $\epsilon_2$ or $\epsilon_3$
we will have respectively $\lambda_1=m$ or $\lambda_1+\lambda_2=m$,
$\lambda_2=m$ or $\lambda_1'=m$, $\lambda_2'=m$ or
$\lambda_1'+\lambda_2'=m$. Counting all these occurrences
involves counting the number of walks $a\in W_{3n}^{(3)}$ which have
a fixed step $\Lambda_p,\Lambda_{p+1}$. These paths are made of two
pieces: 
\item{(i)} the portion $\Lambda_0$, $\Lambda_1$, ..., $\Lambda_p$
which goes from the origin $\Lambda_0=(1,1)$ to $\Lambda_p$ on
$\Pi_+$. There are $C_{\Lambda_p}^{(p)}$ such paths (see \refty).
\item{(ii)} the portion $\Lambda_{p+1}$, 
$\Lambda_{p+2}$, ..., $\Lambda_{3n}=(1,1)$,
which can be thought of as the ``reverse" 
path $\Lambda_0'=\Lambda_{3n}^t=(1,1)$,
$\Lambda_1'=\Lambda_{3n-1}^t$, ..., $\Lambda_{3n-p-1}'=\Lambda_{p+1}^t$
of $3n-p-1$ steps, from the origin
to $\Lambda_{p+1}^t$ (the superscript $t$ means
$(\lambda_1,\lambda_2)^t=(\lambda_2,\lambda_1)$)
obtained by ``reversing" the directions of all the steps, 
namely exchange all
$\epsilon_1\leftrightarrow\epsilon_3$. There are 
$C_{\Lambda_{p+1}^t}^{(3n-p-1)}$ such paths (see \refty).
\par
\noindent{}Hence the number of paths with a specified step
$(\Lambda_p,\Lambda_{p+1})$ is $C_{\Lambda_p}^{(p)}
C_{\Lambda_{p+1}^t}^{(3n-p-1)}$.  
We are now ready to express the total number of occurrences of $\mu_m$
in \detcal, namely
\eqn\powmu{\eqalign{ \alpha_{m,n}^{(3)}~&=~
{1\over 2}\sum_{p,\lambda}
C_{(m,\lambda)}^{(p)} C_{(\lambda,m+1)}^{(3n-p-1)}
+{1\over 2} \sum_{p,\lambda} C_{(\lambda,m-\lambda)}^{(p)}
C_{(m-\lambda,\lambda+1)}^{(3n-p-1)} \cr
&+{1\over 2}\sum_{p,\lambda}
C_{(\lambda,m)}^{(p)} C_{(m+1,\lambda-1)}^{(3n-p-1)}
+{1\over 2} \sum_{p,\lambda} C_{(m+1,\lambda-1)}^{(p)}
C_{(\lambda,m)}^{(3n-p-1)} \cr
&+{1\over 2}\sum_{p,\lambda}
C_{(\lambda,m+1)}^{(p)} C_{(m,\lambda)}^{(3n-p-1)}
+{1\over 2} \sum_{p,\lambda} C_{(\lambda,m+1-\lambda)}^{(p)}
C_{(m-\lambda,\lambda)}^{(3n-p-1)} \cr}}
where the sums extend over $p=0,1,...,3n-1$ and
$\lambda\geq 1$ such that the weights stay in $\Pi_+$,
and each line corresponds to the terms coming from each line of
\weight.
The summations in \powmu\ can be rearranged into 
\eqn\reasum{ \alpha_{m,n}^{(3)}~=~\sum_{p,\lambda} \big[
C_{(m,\lambda)}^{(p)} C_{(\lambda,m+1)}^{(3n-p-1)} 
+C_{(\lambda,m-\lambda)}^{(p)} C_{(m-\lambda,\lambda+1)}^{(3n-p-1)}
+C_{(\lambda,m)}^{(p)} C_{(m+1,\lambda-1)}^{(3n-p-1)} \big]}
It is then a straightforward though tedious exercise in 
combinatorics to prove that
\eqn\protva{ \alpha_{m,n}^{(3)}~=~C_{(m,m)}^{(3n)}-C_{(m+2,m-1)}^{(3n)}
-C_{(m-1,m+2)}^{(3n)}+C_{(m+1,m+1)}^{(3n)}} 
for $m\geq 2$,
by use of the definition \refty\ of the $C$'s in terms of $D$'s.
For $m=1$, we only have
\eqn\protov{ \alpha_{1,n}^{(3)}~=~C_{(2,2)}^{(3)} }

Finally, we write
\eqn\determ{ \Delta_{3n}^{(3)}(\beta)~=~ \prod_{m=1}^n
(\mu_m)^{-\alpha_{m,n}^{(3)}}}
and the result \fordetroi\ follows from the definition of $\mu_m$
\nota, with
$a_{m,n}^{(3)}=\alpha_{m,n}^{(3)}-\alpha_{m+1,n}^{(3)}$, which
amounts to \aleq. 

\newsec{SU(N) meander determinant}

In this section, we present the generalization to $SU(N)$ of
the notion of meander, through pairs of walk diagrams, in
one to one correspondence with reduced elements of a particular
ideal ${\cal I}_{Nn}^{(N)}(\beta)$ of the $SU(N)$ quotient
$H_{Nn}^{(N)}(\beta)$ of the Hecke algebra $H_{Nn}(\beta)$, in
which all antisymmetrizers of order $N+1$ vanish.
The orthonormalization of a basis of this ideal yields a
formula for the corresponding generalized meander determinant.

\subsec{SU(N) walk diagrams}

Let  us 
denote by $\Lambda=\sum \lambda_i\omega_i=
(\lambda_1,...,\lambda_{N-1})$, $\lambda_i\in \IZ$, the elements of the
weight lattice of the $sl(N)$ algebra, generated
by the fundamental weights $\omega_1$, $\omega_2$, ..., $\omega_{N-1}$
in $\IR^{N-1}$, with the scalar products 
\eqn\scapro{ \omega_i \cdot \omega_j ~=~ {i(N-j) \over N} }
for $1 \leq i\leq j\leq N-1$. 
The Weyl chamber $P_+\subset P$ is defined
as the set of weights
\eqn\chambre{ P_+~=~\{ (\lambda_1,...,\lambda_{N-1})
\ {\rm such}\ {\rm that}\ \lambda_i\geq 1\ {\rm for}\ {\rm all}\ i\}}
The Weyl group of $sl(N)$ is the group generated by the reflections
$s_i$ w.r.t. the walls of the Weyl chamber, i.e., the 
hyperplanes $\lambda_i=0$. It is
isomorphic to the permutation group $S_N$ of $N$ objects. The Weyl
chamber is nothing but the quotient of the weight lattice by the action 
of this group.

The weight lattice and Weyl chamber are made into simplices respectively
denoted by $\Pi$ and $\Pi_+$, by the adjunction of oriented links
between the weights, along the vectors
\eqn\vecep{ \epsilon_i=\omega_i-\omega_{i-1} , \ \ i=2,3,...,N-1 }
and $\epsilon_1=\omega_1$, $\epsilon_N=-\omega_{N-1}$, with the
property that $\sum\epsilon_i=0$.
Let us denote by $\rho=(1,1,...,1)$ the origin (apex) of the Weyl chamber
$P_+$.

A $SU(N)$ walk diagram of order $Nn$ is a closed path of $Nn$ steps
on $\Pi_+$ starting and ending at $\rho$. It is uniquely determined
by a sequence $\Lambda_0=\rho$, $\Lambda_1$, ..., $\Lambda_{Nn-1}$,
$\Lambda_{Nn}=\rho$ of weights in $P_+$, satisfying
\eqn\satis{ \Lambda_{i}-\Lambda_{i-1} \in 
\{\epsilon_1,\epsilon_2,...,\epsilon_N\}}
for all $i=1,2,...,Nn$. As before, the index $i$ in $\Lambda_i$ is referred
to as the position of the weight $\Lambda_i$ in the walk diagram.
The set of $SU(N)$ walk diagrams of order $Nn$ is denoted by
$W_{Nn}^{(N)}$.
We can still represent the $SU(N)$ walk diagrams pictorially
in the plane, by
replacing each step $\epsilon_i$ 
by an edge of unit length, making an angle of
${\pi \over 2N}(N-2i+1)$ with the horizontal axis,
and connecting the successive edges of each walk diagram.
For illustration the $SU(4)$ edges read
\eqn\defedge{ 
\figbox{4.cm}{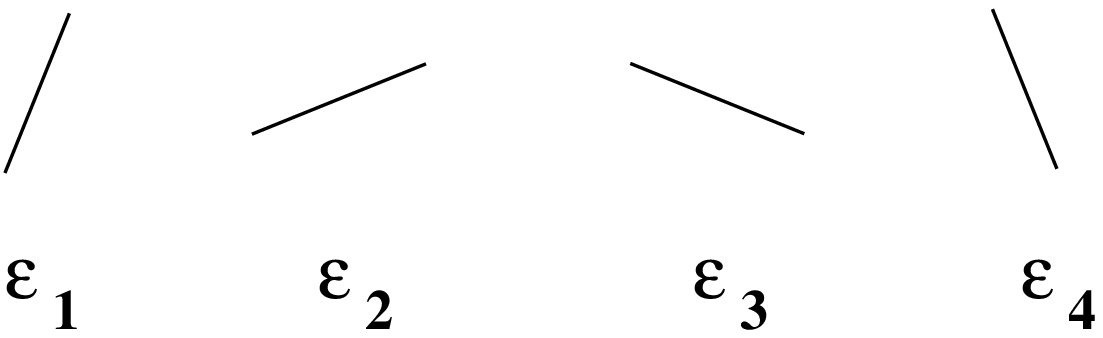}}
We also define the fundamental $SU(N)$ walk diagram $a_0^{(N)}$ with
the successive weights $\Lambda_{i+Nj}-\Lambda_{i+Nj-1}=\epsilon_i$, 
for $i=1,2,...,N$ and $j=0,1,...,n-1$.

To count the number $c_{Nn}^{(N)}$ of $SU(N)$ walk diagrams of order $Nn$, 
let us first compute the number $D_{\Lambda}^{(M)}$
of paths of $M$ steps on 
$\Pi$ from the origin $(0,0,...,0)$ to a fixed weight $\Lambda$.
Assume this path is made of $n_1$ steps $\epsilon_1$, 
$n_2$ steps $\epsilon_2$, ..., $n_N$ steps $\epsilon_N$, 
then we must have $\Lambda~=~\sum_{1\leq i\leq N}n_i\epsilon_i$, and
$n_1+n_2+...+n_N=M$. This is easily inverted into
\eqn\inverty{ n_i~=~ \epsilon_i \cdot \Lambda +{M\over N} }
for $i=1,2,...,N$. The $n_i$ are integers only if 
$M-\sum i\lambda_i=0$ mod $N$, otherwise there is no path of $M$
steps from $(0,0,...,0)$ to $\Lambda$, and $D_{\Lambda}^{(M)}=0$.
The paths are then obtained by arbitrarily choosing the 
$n_i$ steps $\epsilon_i$, resulting in
\eqn\fordh{ D_{\Lambda}^{(M)}~=~ {M! \over \prod_{i=1}^N ({M\over N}
+\epsilon_i\cdot \Lambda)! } }

The number $c_{Nn}^{(N)}$ of $SU(N)$ walk diagrams of $Nn$ steps
is now obtained by subtracting from the number of closed
paths on $\Pi$ from the origin $\rho$ to itself, the  
number of
paths which cross the walls of the Weyl chamber, namely the
hyperplanes $\lambda_i=0$, $i=1,2,...,N-1$.
This is done by the following alternate sum over the images of the 
origin $\rho$ of $\Pi_+$ under the action of the Weyl group
of $sl(N)$ (isomorphic to $S_N$), 
generated by the reflections w.r.t. the walls of the 
Weyl chamber:
\eqn\waldian{\eqalign{
c_{Nn}^{(N)}~&=~\sum_{\sigma\in S_N} (-1)^{l(\sigma)}
D_{\rho-\sigma(\rho)}^{(Nn)}\cr
&=~(Nn)!\prod_{i=0}^{N-1} {i! \over (n+i)!} \cr}}
which gives a natural $SU(N)$ generalization of the Catalan numbers
\catal.

$$\vbox{\font\bidon=cmr8 \def\bidondon{\bidon} \bidondon
\offinterlineskip
\halign{\tv \quad # \tv &
\hfill \ # & \hfill # & \hfill # &  \hfill #
& \hfill # &  \hfill # \tv \cr
\noalign{\hrule}
\tvi  $\scriptstyle N\backslash n$ & 1 \hfill & 2 \hfill
& 3 \hfill & 4 \hfill & 5 \hfill
& 6 \hfill \cr
\noalign{\hrule}
\tvi 2 & 1 & 2 & 5 & 14 & 42 & 132  \cr
\tvi 3 & 1 & 5 & 42 & 462 & 6006  & 87516 \cr
\tvi 4 & 1 & 14 & 462 & 24024 &  1662804  & 140229804 \cr
\tvi 5 & 1 & 42  & 6006  &  1662804  & 701149020  & 396499770810 \cr
\tvi 6 & 1 & 132  & 87516  & 140229804  & 396499770810  & 
1671643033734960  \cr
\noalign{\hrule}
}}$$
\leg{{\bf Table V:} The numbers $c_{Nn}^{(N)}$
of $SU(N)$ walk diagrams of order $Nn$, with $1\leq n,N\leq 6$.
The symmetry $N\leftrightarrow n$ will be interpreted later as some
general duality property.}

Similarly, the number $C_\Lambda^{(M)}$ of paths of $M$ steps
on $\Pi_+$ from the origin $\rho$ to a given weight 
$\Lambda$ is obtained by
subtracting from the corresponding number
of paths on $\Pi$, $D_{\rho,\Lambda}^{(M)}\equiv D_{\Lambda-\rho}^{(M)}$,
the number of those which cross the walls of the Weyl chamber, namely
\eqn\coefn{\eqalign{C_{\Lambda}^{(M)}~&=~\sum_{\sigma \in S_N}
(-1)^{l(\sigma)} D_{\Lambda-\sigma(\rho)}^{(M)} \cr
&=~M! {\prod_{1\leq i<j\leq N} 
\big( (\epsilon_{i}-\epsilon_{j})\cdot \Lambda\big)!
\over \prod_{i=0}^{N-1} 
\big({M\over N}+\epsilon_i\cdot (\Lambda-\rho)+i\big)!} \cr}} 
where the second line follows from the celebrated 
Weyl character formula.
Again, the number $C_\Lambda^{(M)}$ vanishes unless 
$M-\sum i (\lambda_i-1)=0$ mod $N$.

\subsec{Hecke algebra SU(N) quotient and ideal}

We now concentrate on the quotient $H_{Nn}^{(N)}(\beta)$ of
the Hecke algebra $H_{Nn}(\beta)$ \hec, by the generalized
Young antisymmetrizers of order $N+1$, namely defined
by the conditions \hec, supplemented by
\eqn\youN{ Y(e_i,e_{i+1},...,e_{i+N-1})~=~0}
for $i=1,2,...,N(n-1)$.
We now consider the left 
ideal ${\cal I}_{Nn}^{(N)}(\beta)$, generated by the element
\eqn\genid{ Y_{Nn}^{(N)}~=~ \prod_{i=0}^{n-1}
Y(e_{iN+1},e_{iN+2},...,e_{iN+N-1}) }

There is a one-to-one correspondence between the $SU(N)$ walk
diagrams of order $Nn$ and the reduced elements of 
${\cal I}_{Nn}^{(N)}(\beta)$. To properly construct it, we
first need to express the $SU(N)$ walk diagrams as the
results of successive box additions on the fundamental diagram
$a_0^{(N)}$.
Given a walk diagram $a\in W_{Nn}^{(N)}$, the process of box
addition at position $i$ on $a$, producing a diagram $b=a+\diamond_i$,
is allowed only if $a$ has a minimum at $i$, 
namely $N\geq r>s\geq 1$, if 
$\Lambda_{i+1}-\Lambda_i=\epsilon_s$ and 
$\Lambda_i-\Lambda_{i-1}=\epsilon_r$.
The box addition amounts to replacing $\Lambda_i\to \Lambda_i+\epsilon_s
-\epsilon_r$, i.e., exchanging the two steps $\epsilon_r$ and $\epsilon_s$
in the corresponding path on $\Pi_+$. In the above 
pictorial representation, a box addition amounts to adding to $a$ a
parallelogram (the ``box"), with edges corresponding to the vectors 
$\epsilon_r$ and $\epsilon_s$. This gives rise to $N(N-1)/2$ different
types of boxes.
It is clear that any walk $a\in W_{Nn}^{(N)}$ can be obtained from the
fundamental one $a_0^{(N)}$ by successive box additions. 
As in the $SU(3)$ case, the box decomposition of a given walk $a$ is not
unique, due to all possible hexagon ambiguities. Indeed, for any three
integers $N\geq i>j>k\geq 1$, there are two possibilities to change
the succession of steps $(\epsilon_i,\epsilon_j,\epsilon_k)$ into
$(\epsilon_k,\epsilon_j,\epsilon_i)$ by three successive box additions:
\eqn\hexambi{ \figbox{2.4cm}{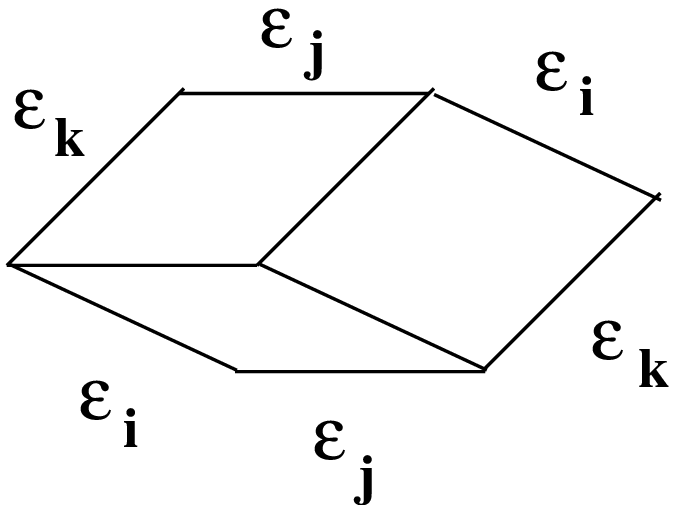} \quad {\rm or} \quad 
\figbox{2.4cm}{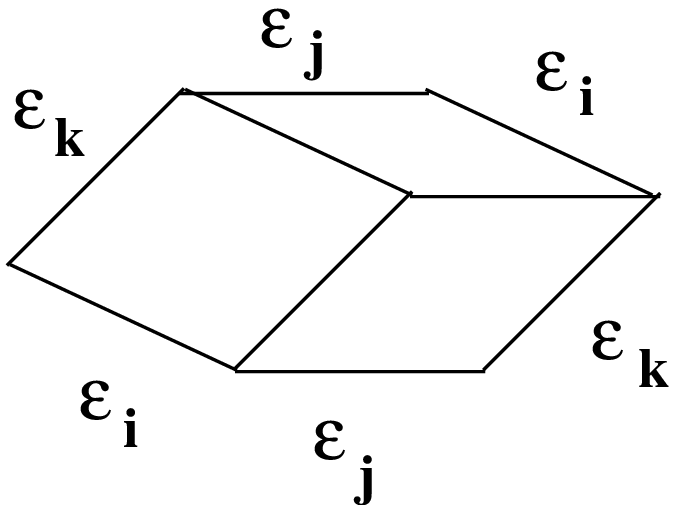} }
To resolve these ambiguities, we forbide all the box additions of the form
\eqn\foradbo{\figbox{1.6cm}{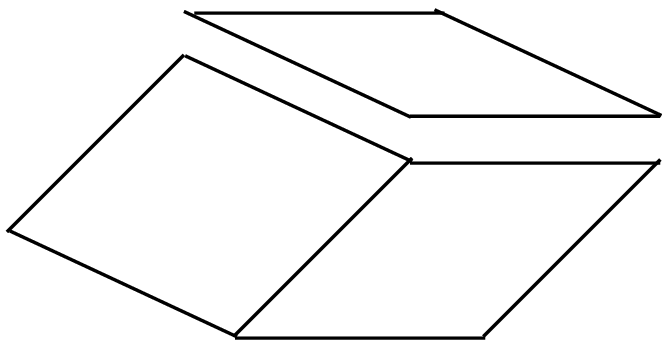} }
for all $N\geq i>j>k\geq 1$. With this last rule, each walk 
$a\in W_{Nn}^{(N)}$ has a unique box decomposition, represented as the
set of boxes inbetween $a_0^{(N)}$ and $a$. 
\fig{A sample $SU(4)$ walk diagram $a$ of order $4\times 3=12$
is represented in thick line. 
It is made of a succession of steps of the form \defedge.
We have indicated its box decomposition in thin lines, 
leading from the
fundamental diagram $a_0^{(4)}$ of order $12$ to $a$,
after a total of $|a|=10$ box additions.}{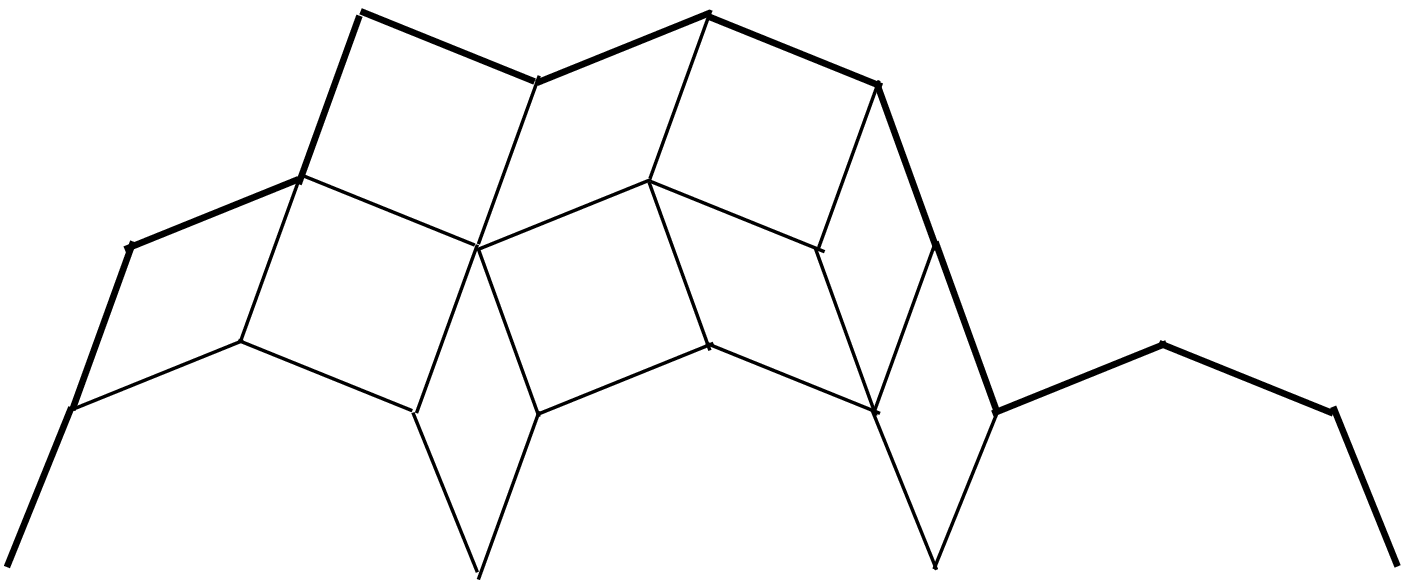}{6.cm}
\figlabel\samfour
\noindent
For illustration, we 
display in Fig.\samfour\ a sample $SU(4)$ walk 
diagram of order $12$, together with
its box decomposition.

We can now construct the map $\varphi$ from $W_{Nn}^{(N)}$ to 
the set of reduced elements of ${\cal I}_{Nn}^{(N)}(\beta)$, through
\eqn\iniphi{ \varphi(a_0^{(N)})~=~ Y_{Nn}^{(N)} }
with $Y_{Nn}^{(N)}$ as in \genid, and the recursion on box
additions, for any $a\in W_{Nn}^{(N)}$
\eqn\rebN{ \varphi(a+\diamond_i)~=~ e_i \ \varphi(a) }
This produces exactly once all the reduced elements of 
${\cal I}_{Nn}^{(N)}(\beta)$. For illustration,
the walk diagram of Fig.\samfour, 
has the following image under $\varphi$:
$e_4e_5e_6e_2e_3e_5e_6e_7e_4e_8
Y(e_1,e_2,e_3)Y(e_5,e_6,e_7) Y(e_9,e_{10},e_{11})$, as a result of $10$
box additions on the fundamental diagram.      
As before, we
introduce the basis 1, with elements
\eqn\elebaN{ (a)_1~=~ 
(\alpha_N)^{-n/2}
(\gamma_N)^n \ \varphi(a)}
where $\alpha_N$ and $\gamma_N$ are defined in  \procons.
The normalization, somewhat arbitrary, is chosen for reasons which will
become clear later.

The $SU(N)$ meanders are defined as pairs of walks $(a,b)\in W_{Nn}^{(N)}$, 
or equivalently of elements of this basis 1.
To the latter, we attach the quantity $\big((a)_1,(b)_1 \big)$,
where the scalar product is attached to the Markov trace ${\rm Tr}$
on $H_{Nn}^{(N)}(\beta)$,
defined by the normalization ${\rm Tr}(1)=(U_{N-1})^{Nn}$ and the recursion
\marko, with $\eta=\mu_{N-1}=U_{N-2}(\beta)/U_{N-1}(\beta)$.
This leads to the $c_{Nn}^{(N)}\times c_{Nn}^{(N)}$ Gram matrix
${\cal G}_{Nn}^{(N)}(\beta)$, with entries
\eqn\graN{ \big[ {\cal G}_{Nn}^{(N)}(\beta)\big]_{a,b}~=~
\big((a)_1,(b)_1\big)}
for $a,b\in W_{Nn}^{(N)}$.

\subsec{SU(N) determinant}
\par
The main result of this section is the following formula for the determinant 
$\Delta_{Nn}^{(N)}(\beta)$ of the Gram matrix \graN:
\eqn\sundet{\Delta_{Nn}^{(N)}~=~\prod_{m=1}^{n+N-2}
\big[ U_m(\beta)\big]^{a_{m,n}^{(N)}} }
where $a_{m,n}^{(N)}$ are some integers, defined as follows.
\fig{The set of $12$ vectors forming the difference operator
defining $a_{m,n}^{(N)}$ in terms of $C_\Lambda^{(Nn)}$, for
$N=4$. We have indicated the vectors
$\epsilon_{i,j}\equiv\epsilon_i-\epsilon_j$ linking the dots,
representing $(m-1,1,m-1)+u_k+v_l$ on the left half ($\Delta_4$ operator),
and  $(m+2,1,m+2)-{\bar u}_k-{\bar v}_l$ on the right half (${\bar
\Delta}_4$ operator). 
The terms with a filled black circle come with a $+$,
those with an empty circle with a $-$ in the final difference 
operator.}{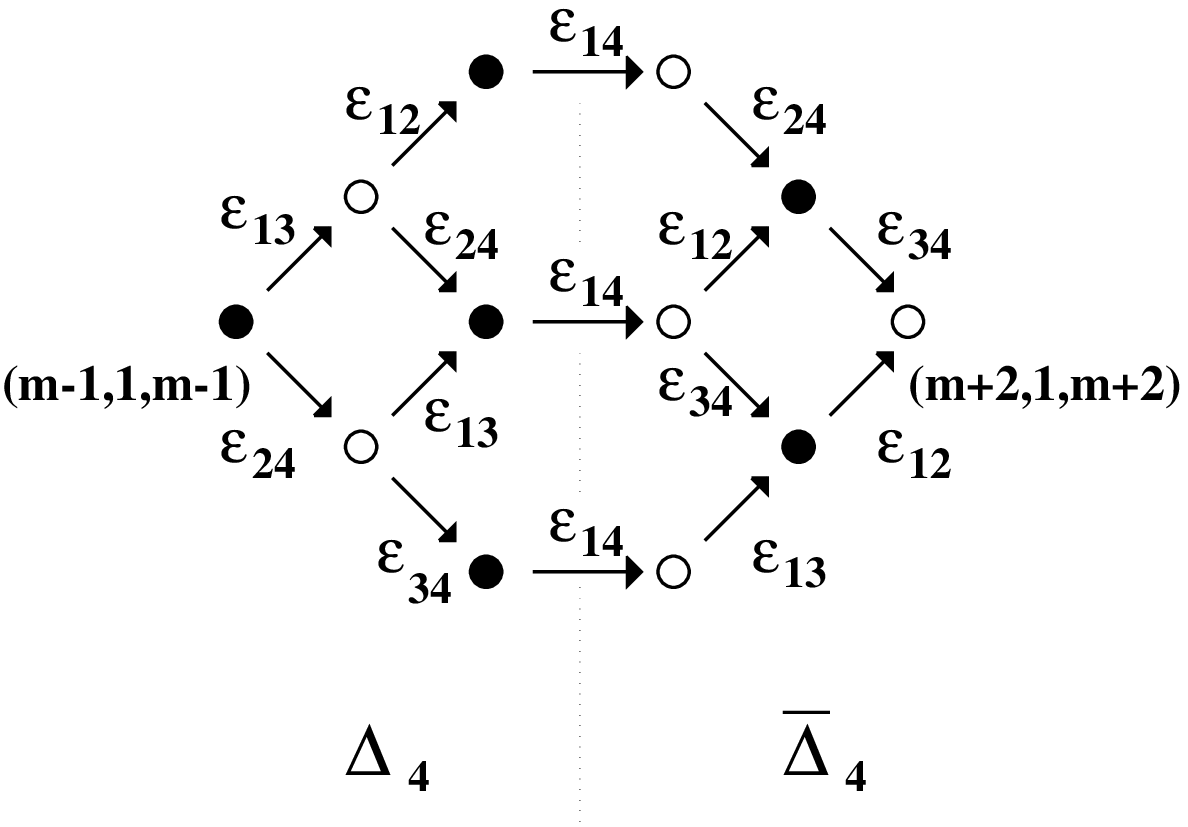}{6.cm}
\figlabel\differ
We first introduce 
the vectors $u_0=v_0={\bar u}_0={\bar v}_0=0$, and 
\eqn\folvec{\eqalign{
u_j~&=~ j \epsilon_1 -(\epsilon_j+\epsilon_{j+1}+...+\epsilon_{N-1})\cr
v_j~&=~ -j\epsilon_N +(\epsilon_2+\epsilon_3+...+\epsilon_{j+1})\cr
{\bar u}_j~&=~j(\epsilon_1-\epsilon_N)-u_{N-1-j}\cr
{\bar v}_j~&=~j(\epsilon_1 -\epsilon_N)-v_{N-1-j}\cr}}
for $j=1,2,...,N-2$ (see Fig.\differ\ for an illustration in the
case $N=4$).
We define the following difference operators $\Delta_N$, and
${\bar \Delta}_N$
acting on any function
$f(\alpha)$, of $\alpha\in P$ by the alternate sums
\eqn\difop{\eqalign{
[\Delta_N f](\alpha)~&=~\sum_{i,j\geq 0\atop
i+j\leq N-2} (-1)^{i+j}\ f(\alpha +(u_i+v_j))\cr 
[{\bar \Delta}_N f](\alpha)~&=~ \sum_{i,j\geq 0 \atop
i+j\leq N-2} (-1)^{i+j}\ f(\alpha-({\bar u}_i+{\bar v}_j))\cr}}
We also define $\Delta_N^{*}$ as the same expression as 
for $\Delta_N$, except that 
the point $i=j=0$ is excluded from the sum \difop.
Now, we use the function $f(\Lambda)=C_\Lambda^{(Nn)}$,
$\Lambda\in P_+$, and $C_\Lambda^{(Nn)}$ as in \coefn, 
to write the integers $a_{m,n}^{(N)}$ as
\eqn\writf{  a_{m,n}^{(N)}~=~ 
\Delta_N f
\big(\rho+(m-N+2)(\epsilon_1-\epsilon_N)\big) 
-{\bar\Delta}_N f\big(\rho+(m+1)(\epsilon_1-\epsilon_N)\big) }
for any integer $m\neq N-2$ and
\eqn\writo{ a_{N-2,n}^{(N)}~=~
\Delta_N^{*}f
\big(\rho\big) 
-{\bar\Delta}_N f\big(\rho+(N-1)(\epsilon_1-\epsilon_N)\big)}
We have already noticed in the cases $N=2$ and $3$ that the 
numbers $a_{m,n}^{(N)}$  take the form of an
alternate sum over the Weyl group. Here we see that the correct
generalization \writf\ is not an alternate sum over the Weyl group,
but only over a set of $N(N-1)$ shifted weights, represented in 
Fig.\differ\ for $N=4$.
In this latter case, the corresponding $12$ term-relation reads
\eqn\twelve{\eqalign{
a_{m,n}^{(4)}~&=~C_{(m-1,1,m-1)}^{(4n)}-C_{(m,2,m-2)}^{(4n)}
-C_{(m-2,2,m)}^{(4n)}+C_{(m+2,1,m-2)}^{(4n)}\cr
&+C_{(m-2,1,m+2)}^{(4n)}+C_{(m-1,3,m-1)}^{(4n)}
-C_{(m+3,1,m-1)}^{(4n)}-C_{(m-1,1,m+3)}^{(4n)} \cr
&-C_{(m,3,m)}^{(4n)}+C_{(m+2,2,m)}^{(4n)}+C_{(m,2,m+2)}^{(4n)}
-C_{(m+2,1,m+2)}^{(4n)} \cr}}
for $m\neq 2$, and
\eqn\otre{a_{2,n}^{(4)}~=~C_{(1,3,1)}^{(4n)}-C_{(5,1,1)}^{(4n)}
-C_{(1,1,5)}^{(4n)}-C_{(2,3,2)}^{(4n)}+C_{(4,2,2)}^{(4n)}+
C_{(2,2,4)}^{(4n)}-C_{(4,1,4)}^{(4n)} }
The first few values for the integers $a_{m,n}^{(N)}$
are given in Tables VI-a,b
for the cases $N=4,5$ respectively.

$$\vbox{\font\bidon=cmr8 \def\bidondon{\bidon} \bidondon
\offinterlineskip
\halign{\tv \quad # \tv &
\hfill \ # & \hfill # & \hfill # &  \hfill #
& \hfill # &  \hfill #
\tv \cr
\noalign{\hrule}
\tvi  $\scriptstyle m\backslash n$& 1 \hfill & 2 \hfill
& 3 \hfill & 4 \hfill & 5 \hfill
& 6 \hfill \cr
\noalign{\hrule}
\tvi 1 & 1 & 20 & 627 & 24024 & 831402  & -8498776  \cr
\tvi 2 & 1 & 15 & 572 & 36036 & 2922504  & 274085526 \cr
\tvi 3 & 1 & 22 & 880 & 48048 & 3375996  & 291900268 \cr
\tvi 4 &   & 13 & 550 & 36036 & 2910876  & 265913626 \cr
\tvi 5 &   &    & 341 & 24024 & 1951566  & 192203088 \cr
\tvi 6 &   &    &     & 12012 & 1372104  & 139085738 \cr
\tvi 7 &   &    &     &       &  492252  & 85314636  \cr
\tvi 8 &   &    &     &       &          & 22064130  \cr
\noalign{\hrule}
}}$$
\leg{{\bf Table VI-a:} The powers $a_{m,n}^{(4)}$ of $U_m$ in
the $SU(4)$ meander determinant $\Delta_{4n}^{(4)}(\beta)$,
for $n=1,2,...,6$.}

$$\vbox{\font\bidon=cmr8 \def\bidondon{\bidon} \bidondon
\offinterlineskip
\halign{\tv \quad # \tv &
\hfill \ # & \hfill # & \hfill # 
& \hfill # &  \hfill # \tv \cr
\noalign{\hrule}
\tvi  $\scriptstyle m\backslash n$ & 1 \hfill & 2 \hfill
& 3 \hfill & 4 \hfill & 5 \hfill \cr
\noalign{\hrule}
\tvi 1 & 1 & 69 & 10582 & 2494206 & 701149020    \cr
\tvi 2 & 1 & 44 & 6435  & 2065908 & 1051723530   \cr
\tvi 3 & 1 & 58 & 10712 & 3275220 & 1402298040   \cr
\tvi 4 & 1 & 76 & 12311 & 3740340 & 1752872550   \cr
\tvi 5 &   & 41 &  8736 & 3036846 & 1402298040   \cr
\tvi 6 &   &    &  5278 & 1953504 & 1051723530 \cr
\tvi 7 &   &    &       & 1170552 & 701149020  \cr
\tvi 8 &   &    &       &         & 350574510  \cr
\noalign{\hrule}
}}$$
\leg{{\bf Table VI-b:} The powers $a_{m,n}^{(5)}$ of $U_m$ in
the $SU(5)$ meander determinant $\Delta_{5n}^{(5)}(\beta)$,
for $n=1,2,...,5$.}

The formula \sundet\ is proved by direct orthogonalization of the basis 1,
namely the construction of a basis 2, with elements
\eqn\batof{ (a)_2~=~ \sum_{b\in W_{Nn}^{(N)}\atop b \subset a} 
P_{a,b} (b)_1 }
where the inclusion $b\subset a$ means that 
$a$ can be obtained from $b$ by box additions, and such that
$\big( (a)_2,(b)_2\big)=\delta_{a,b}$.
The basis $2$ is constructed as follows. We start with
\eqn\starN{ (a_0^{(N)})_2~=~(\alpha_N)^{n/2} (a)_1 }
with $\alpha_N$ as in \procons.
This element has norm $1$, due to \idmpo, and the property
\eqn\trN{ {\rm Tr}\big(Y(e_1,...,e_{N-1})\big)~=~\gamma_N^{-1}  }
where the trace is taken over $H_N^{(N)}(\beta)$.
This is readily proved by use  
of the recursion of the $Y$'s \recyou. 
We finally get ${\rm Tr}\big((\gamma_N Y)^2\big)=1$.

The other basis 2 elements are defined through the recursion relation,
for any $a\in W_{Nn}^{(N)}$
\eqn\recurN{ (a+\diamond_{i,m})_2~=~\sqrt{\mu_{m+1}\over \mu_m}
(e_i -\mu_m) \ (a)_2 }
where $i$ denotes as usual the position of the box addition, and
$m$ stands for the ``height" of the box addition, defined as
\eqn\highN{m~=~\Lambda_i \cdot (\Lambda_{i+1}+\Lambda_{i-1}-2\Lambda_i)}
in terms of the weights $\Lambda_{i-1}$, $\Lambda_i$ and $\Lambda_{i+1}$ of
respective positions $i-1$, $i$ and $i+1$ on $a$.
The basis 2 coincides with the restriction to ${\cal I}_{Nn}^{(N)}(\beta)$
of the semi-normal basis of \DJ\ \MUR, and as such satisfies the
condition
\eqn\condiN{ (a)_2^t (b)_2~=~0 \ \ {\rm unless} \ \ a~=~b }
Assuming \condiN,
with exactly the same reasoning as in \simuop, it is easy to check that 
the normalization prefactor in \recurN\ ensures that all the basis 2 elements
have norm 1.
This change of basis resolves the hexagon ambiguities, in the form of a 
straightforward generalization of \twoeq, to all the possible hexagons
\hexambi.

As before, the $SU(N)$ meander determinant reads
\eqn\detNme{ \Delta_{Nn}^{(N)}(\beta)~=~ \prod_{a \in W_{Nn}^{(N)}}
P_{a,a}^{-2} }
with $P$ as in \batof.
In turn, we have 
\eqn\inipN{ P_{a_0^{(N)},a_0^{(N)}}^2~=~
(\prod_{i=1}^{N-1}(\mu_i)^{N-i})^n }
as a direct consequence of \batof\ and \starN\ ($(a_0^{(N)})_2$ has norm
1), and the recursion relation
\eqn\recpaaN{ P_{a+\diamond_{i,m},a+\diamond_{i,m}}^2~=~
{\mu_{m+1}\over \mu_m} P_{a,a} }
for all $a\in W_{Nn}^{(N)}$, solved as 
\eqn\decNm{ P_{a,a}^2~=~ (\mu_1^{N-1}\mu_2^{N-2}\cdots \mu_{N-1})^n 
\prod_{{\rm all}\ {\rm boxes}\atop {\rm of } \ a} 
{\mu_{m+1} \over \mu_m} }
As in the $SU(3)$ case, we rearrange the $\mu$ factors into 
edge weights. 
More precisely, to each edge $(\Lambda,\Lambda')$ of $a$, corresponding
to a step $\Lambda'-\Lambda=\epsilon_i$, we attach
the weight 
\eqn\wN{ w(\Lambda,\Lambda')~=~\left( 
\prod_{l=i+1}^N
\mu_{\Lambda\cdot (\epsilon_i-\epsilon_l)} 
\prod_{k=1}^{i-1} \mu_{\Lambda'\cdot (\epsilon_k-\epsilon_i)} 
\right)^{1\over 2} }
in terms of which \decNm\ is rewritten as
\eqn\dNm{ P_{a,a}^2~=~ \prod_{{\rm all}\ {\rm steps}\atop
v \ {\rm of } \ a} w(v) }
This is easily proved by induction on box additions, 
as the r.h.s. of \dNm\ satisfies
both \inipN\ and \recpaaN. Indeed, the box addition
$a \to a+\diamond_{i,m}$ transforms
the sequence of weights of $a$ 
$(\Lambda_{i-1},\Lambda_i,\Lambda_{i+1})$, with
say $\Lambda_{i}-\Lambda_{i-1}=\epsilon_s$ and $\Lambda_{i+1}
-\Lambda_i=\epsilon_r$, into the sequence
$(\Lambda_{i-1},\Lambda_i',\Lambda_{i+1})$ with
$\Lambda_i'-\Lambda_{i-1}=\epsilon_r$; we then check that
the edge weights satisfy
$w(\Lambda_{i-1},\Lambda_i')w(\Lambda_i',\Lambda_{i+1})/
[w(\Lambda_{i-1},\Lambda_i)w(\Lambda_i,\Lambda_{i+1})]
=\mu_{m+1}/\mu_m$, with $m=\Lambda_i\cdot (\epsilon_r-\epsilon_s)$,
hence \recpaaN\ follows.
After substitution of \dNm\ into \detNme, we are left with an
expression of the form
\eqn\forN{ \Delta_{Nn}^{(N)}(\beta)~=~ \prod_{m=1}^{N+n-1} 
\mu_m^{-\alpha_{m,n}^{(N)}}  }

To compute $\alpha_{m,n}^{(N)}$, we have to enumerate the walks
containing edges, whose weight contains a $\mu_m$. According to
the definition \wN, such an edge is of the form
$(\Lambda,\Lambda')$, $\Lambda'-\Lambda=\epsilon_i$, 
with either $\Lambda'\cdot (\epsilon_k -\epsilon_i)=m$  for some $k<i$
or $\Lambda\cdot (\epsilon_i-\epsilon_l)=m$ for some $l>i$.
To proceed, we must count the number of edges of walks of
$W_{Nn}^{(N)}$ with specified
ends $(\Lambda,\Lambda')$, at positions, say $p$ and $p+1$.
This edge cuts the walk $a$ into two portions 
\item{(i)} a path of $p$ steps from the origin $\rho$ to $\Lambda$
on $\Pi_+$. There are $C_\Lambda^{(p)}$ such paths (see \coefn).
\item{(ii)} a path of $Nn-p-1$ steps from $\Lambda'$ to $\rho$ on $\Pi_+$.
Upon reversal of all the orientations of its links
(namely exchanging $\epsilon_i \leftrightarrow \epsilon_{N+1-i}$), 
it can be viewed
as the reversed path of $Nn-p-1$ steps from $\rho$ to $\Lambda'^t$ (where
$(\lambda_1,\lambda_2,...,\lambda_{N-1})^t=
(\lambda_{N-1},\lambda_{N-2},...,\lambda_1)$).
There are $C_{\Lambda'^t}^{(Nn-p-1)}$ such paths (see \coefn).
\par
\noindent{}We may now compute the numbers $\alpha_{m,n}^{(N)}$
of \forN, with the result
\eqn\compuN{ \alpha_{m,n}^{(N)}~=~\sum_{p=0}^{Nn-1}
\sum_{1 \leq i<l\leq N} \sum_{\Lambda \in P_+\atop
\Lambda\cdot (\epsilon_i -\epsilon_l)=m}
C_{\Lambda}^{(p)} C_{(\Lambda+\epsilon_i)^t}^{(Nn-p-1)} }
where we have assembled all the contributions from the weight factors 
$w(\Lambda,\Lambda+\epsilon_i)$ as well as those from
$w(\Lambda'-\epsilon_{N+1-i},\Lambda')$, by noting that
$(\Lambda'+\epsilon_i)^t=\Lambda'-\epsilon_{N+1-i}$.

The final formula \sundet\ follows from the definition \nota\
of $\mu_m$ and 
$a_{m,n}^{(N)}=\alpha_{m,n}^{(N)}-\alpha_{m+1,n}^{(N)}$.
Let us first compute the numbers $\alpha_{m,n}^{(N)}$
\compuN. We can use the expression \coefn\
for the $C$'s in terms of the
$D$'s, which are multinomial coefficients, to evaluate the various
sums in \compuN. This finally leads to the following result.
We will use the definition \difop\ for the difference operator
$\Delta_N$.
Let us also define the vectors $w_0=0$ and
\eqn\vecwec{ w_j~=~\epsilon_2+2 \epsilon_3+...+(j-1)\epsilon_{j-1}
+j(\epsilon_j+\epsilon_{j+1}+...+\epsilon_N)}
for $j=1,2,...,N$.
Then, using the function $f(\Lambda)=C_{\Lambda}^{(Nn)}$, 
the integer $\alpha_{m,n}^{(N)}$ reads
\eqn\inteal{
\alpha_{m,n}^{(N)}~=~ \sum_{p\geq 0\atop
2p\leq N-2} \Delta_{N-2p}f 
(\rho+(m-N+2)(\epsilon_1-\epsilon_N)+w_p) }
for $m\geq N-1$. When $1\leq m\leq N-2$, we simply have to omit
the term $i=j=0$ in $\Delta_N$.
After substitution of \difop,
the alternate sum on the r.h.s. of \inteal\
extends over $N(N^2-1)/6$ terms of the form 
$C_\Lambda^{(Nn)}$, $\Lambda=\rho+(m-N+2)(\epsilon_1-\epsilon_N)+
u_i+v_j+w_p$, $0\leq i,j\leq N-2$, $0\leq 2p \leq N-2$
hence forming a ``pyramid" of weights.
The result \writf\ for  the numbers
$a_{m,n}^{(N)}=\alpha_{m,n}^{(N)}-\alpha_{m+1,n}^{(N)}$
then follows from many cancellations between the pyramids of terms of
\inteal\ pertaining to $m$ and $m+1$, leaving us 
eventually with only $N(N-1)$ terms (see Fig.\differ). 
Eq.\writo\ corresponds
to the omission of the term $i=j=0$ in \inteal\ for $m=1,...,N-2$.

\subsec{Duality}

In this section, we describe a duality relation between the
$SU(N)$ and the $SU(k)$ meanders of same order
$Nk$.   
This results in a duality formula for the 
corresponding meander determinants.
\par
The compact definitions \starN\ and \recurN\ for the basis 2 elements
of ${\cal I}_{Nn}^{(N)}(\beta)$ lead us to a simple formula,
relating the $SU(N)$ meander determinant of order $Nk$ to the $SU(k)$
meander determinant of same order, namely
\eqn\dualit{ \Delta_{Nk}^{(N)}(\beta) \Delta_{kN}^{(k)}(\beta)~=~
(\Phi_{N,k} )^{c_{Nk}^{(N)}} }
where, $\Phi_{N,k}$ is symmetric in $k\leftrightarrow N$, and
for $k\leq N$ 
\eqn\dualN{ \Phi_{N,k}~=~\prod_{m=1}^{k-1} (U_m)^{m+1} 
\prod_{m=k}^{N-1} (U_m)^k \prod_{m=N}^{N+k-2} (U_m)^{N+k-1-m} }
in terms of the Chebishev polynomials \ceby.
Note also that from the definition \waldian, 
\eqn\identcnk{c_{Nk}^{(N)}~=~(Nk)!{\prod_{i=1}^{k-1} i! 
\prod_{i=1}^{N-1} i! \over \prod_{i=1}^{N+k-1} i!}
~=~c_{Nk}^{(k)} }
which makes the r.h.s. of \dualit\ symmetric
under $k\leftrightarrow N$.

The duality formula \dualit\ gives a number of combinatorial identities
relating the numbers $a_{m,k}^{(N)}$ and $a_{m,N}^{(k)}$ \writf\-\writo,
namely that, for $k\leq N$  
\eqn\combirul{ a_{m,k}^{(N)}+a_{m,N}^{(k)}~=~
\left\{ \matrix{ &(m+1)c_{Nk}^{(N)}\hfill & ({\rm if}\ m<k) \cr
&k c_{Nk}^{(N)}\hfill & ({\rm if}\ k\leq m<N) \cr
&(N+k-m-1) c_{Nk}^{(N)} \hfill & ({\rm if }\ N\leq m <N+k-1) \cr}\right. }
including the cases \writo\ when $m=N-2$ or $k-2$.

As an example, let us take $N=3$ and $k=2$, in which cases we have
(see Tables II and IV)
\eqn\exdeter{\eqalign{
\Delta_{6}^{(3)}~&=~U_1^6 U_2^6 U_3^4 \cr
\Delta_{6}^{(2)}~&=~U_1^4 U_2^4 U_3 \cr}}
respectively from \detntoex\ and \detforme, with 
$C_3^{(6)}={6 \choose 2}-{6 \choose 1}=9$, 
$C_5^{(6)}={6 \choose 1}-1=5$, $C_7^{(6)}=1$. We check that
\eqn\cheN{ 
\Delta_{6}^{(3)}
\Delta_{6}^{(2)}~=~(U_1^2 U_2^2 U_3)^5}
which amounts to \dualit\-\dualN, with $c_6^{(3)}=c_6^{(2)}=5$ and
$\Phi_{3,2}=U_1^2U_2^2U_3$.
More generally, we can check the above duality relation on the various
Tables II, IV and VI-a,b, by using also the Table V for
the numbers $c_{Nn}^{(N)}$.

A simple consequence of \dualit\ is that the ``self-dual"
determinants, with $k=N$,  read
\eqn\selfdu{\eqalign{ \Delta_{N^2}^{(N)}(\beta)~&=~
(\Phi_{N,N})^{c_{N^2}^{(N)}/2} \cr
&=~(U_1^2 U_2^3 ...U_{N-2}^{N-1} U_{N-1}^N
U_N^{N-1}...U_{2N-3}^2 U_{2N-2})^{c_{N^2}^{(N)}/2} \cr}}
which is considerably simpler than \sundet\ \writf.
This is readily checked for $N=2,3,4,5$ on Tables II, IV and VI-a,b.

The duality formula \dualit\ is a consequence of the existence
of a duality map $\delta$ between the basis 2 elements of
${\cal I}_{Nk}^{(N)}(\beta)$ and ${\cal I}_{Nk}^{(k)}(\beta)$, or
equivalently between their labels $W_{Nk}^{(N)}$ and $W_{Nk}^{(k)}$.
The map $\delta$ is defined as follows. First we need to define
the {\it maximal} walk diagram $a_{max}^{(k)}\in W_{Nk}^{(k)}$, 
as the walk with
$N$ steps $\epsilon_1$, followed by $N$ steps $\epsilon_2$, ...,
followed by $N$ steps $\epsilon_k$. 
In other words, the weights of this walk are 
\eqn\weimax{ \Lambda_{Ni+j}~=~N(\epsilon_1+\epsilon_2+...+\epsilon_i)
+j\epsilon_{i+1} }
for $i=0,1,...,k-1$ and $j=1,2,...,N$.
This walk is maximal w.r.t. box
additions, as it has no minimum, hence no extra box can be added to it.
We also need to define the concept of box subtraction for elements
of the basis 2: 
we will say that $(b)_2$ is the result of a box subtraction
at position $i$ and height $m$ on $(a)_2$,
and write that $(b)_2=(a-\diamond_{i,m})_2$,
if $(a)_2=(b+\diamond_{i,m})_2$ is the result of a box addition
at position $i$ and height $m$ on $b$ (c.f. \recurN).
We will use the same terminology for the corresponding walk diagrams.

The duality map $\delta\ :\ W_{Nk}^{(N)}\to W_{Nk}^{(k)}$ is defined by
\eqn\delini{ \delta(a_0^{(N)})~=~ a_{max}^{(k)} }
and the recursion relation
\eqn\recdelta{ \delta(a+\diamond_{i,m})~=~\delta(a)-\diamond_{i,m} }
In other words, the recursion adds successive boxes
on $a_0^{(N)}$ which it subtracts accordingly from $a_{max}^{(k)}$.
To prove that $\delta$ is well defined, we must simply check
that each minimum on $a$ is a ``maximum" on $\delta(a)$, i.e., 
a position at which a box can be subtracted.
This is clear on $a_0^{(N)}$ and its image, as the $k-1$ minima of 
$a_0^{(N)}$ lie at positions $i=N,2N,...,(k-1)N$, equal to the
positions of the maxima on $a_{max}^{(k)}$, namely the transitions
between the steps $\epsilon_i\to \epsilon_{i+1}$, $i=1,2,...,k-1$.
The recursion then makes it clear that whenever a minimum is created
on $a$ by a neighbouring 
box addition, the corresponding box subtraction on $b$
creates a maximum.  Moreover, as 
$|W_{Nk}^{(N)}|=c_{Nk}^{(N)}=c_{Nk}^{(k)}=|W_{Nk}^{(k)}|$ 
\identcnk, $\delta$ is a bijection.

The computation of the determinants $\Delta_{Nk}^{(N)}(\beta)$ and
$\Delta_{Nk}^{(k)}(\beta)$ involves a product over the quantities
$P_{a,a}^{-2}$ defined by \inipN\ \recpaaN. By a slight abuse of notation, 
we will denote indifferently by $P_{a,a}$ the matrix elements for
both $SU(N)$ and $SU(k)$ cases, simply distinguished
by the fact that $a\in W_{Nk}^{(N)}$ or $W_{NK}^{(k)}$.
Let us prove that, for all $a\in W_{Nk}^{(N)}$ 
\eqn\eqapn{ P_{a,a}^2 P_{\delta(a),\delta(a)}^2 ~=~ 
P_{a_0^{(N)},a_0^{(N)}}^2 P_{a_{max}^{(k)},a_{max}^{(k)}}^2 }
This is readily done by induction on box additions on $a$,
as
\eqn\boxasub{\eqalign{
P_{a+\diamond_{i,m},a+\diamond_{i,m}}^2 
P_{\delta(a+\diamond_{i,m}),\delta(a+\diamond_{i,m})}^2~&=~
P_{a+\diamond_{i,m},a+\diamond_{i,m}}^2 
P_{\delta(a)-\diamond_{i,m},\delta(a)-\diamond_{i,m}}^2\cr
&=~{\mu_{m+1} \over \mu_m} P_{a,a}^2 {\mu_m \over \mu_{m+1}} 
P_{\delta(a),\delta(a)}^2 \cr
&=~P_{a,a}^2 P_{\delta(a),\delta(a)}^2 \cr}}
where we have successively used the recursive definition 
\recdelta\ of $\delta$ and the recursion \recpaaN\ for both 
$P_{a+\diamond,a+\diamond}^2$ and $P_{b+\diamond,b+\diamond}^2$,
with $b=\delta(a)-\diamond$. Eq.\eqapn\ follows.

Therefore, the product of meander determinants
reads
\eqn\promede{\eqalign{ 
\Delta_{Nk}^{(N)}(\beta) \Delta_{Nk}^{(k)}(\beta)
~&=~\prod_{a \in W_{Nk}^{(N)}} P_{a,a}^{-2} \prod_{b\in W_{Nk}^{(k)}}
P_{b,b}^{-2} \cr
&=~\prod_{a \in W_{Nk}^{(N)}} P_{a,a}^{-2} P_{\delta(a),\delta(a)}^{-2}
\cr
&=~\big( P_{a_0^{(N)},a_0^{(N)}}^{-2} P_{a_{max}^{(k)},a_{max}^{(k)}}^{-2}
\big)^{c_{Nk}^{(N)}} \cr}}
as $|W_{Nk}^{(N)}|=c_{Nk}^{(N)}$. The formula \dualit\ follows from
\promede, with
$\Phi_{N,k}^{-1}=P_{a_0^{(N)},a_0^{(N)}}^{2} 
P_{a_{max}^{(k)},a_{max}^{(k)}}^{2}$.
The first factor \inipN\ is known. The second reads, from \dNm\
\eqn\secter{ 
P_{a_{max}^{(k)},a_{max}^{(k)}}^2~=~
\prod_{{\rm all}\ {\rm steps}\atop
v \ {\rm of}\ a_{max}^{(k)}} w(v) }
with the weights $w$ as in \wN\ for $N\to k$ and $n\to N$, and the steps
as in \weimax.
Assembling all the powers of $\mu$, we find
\eqn\finN{ 
P_{a_{max}^{(k)},a_{max}^{(k)}}^2~=~
\prod_{j=1}^N \prod_{i=j}^{j+k-2} (\mu_i)^{k+j+1-i} }
hence finally, for $k\leq N$
\eqn\phifin{\eqalign{ \Phi_{N,k}~&=~\prod_{m=1}^{N-1} (\mu_m)^{-k(N-m)}
\prod_{j=1}^N \prod_{i=j}^{j+k-2} (\mu_i)^{i-(k+j+1)}\cr 
&=~ \prod_{m=1}^{k-1} (\mu_m)^{{m(m+1)\over 2}-Nk} 
\prod_{m=k}^{N} (\mu_m)^{k(m-N-{k-1\over 2})} 
\prod_{m=N+1}^{N+m-2} (\mu_m)^{(m-k-N)(k+N-m-1)\over 2} \cr}}
Using the definition \nota, this is easily translated into the
final result \dualN.

\subsec{Duality and Young tableaux}

This duality is yet another manifestation of the level-rank duality
of the affine Lie algebras $\widehat{sl(n)}_k\leftrightarrow
\widehat{sl(k)}_n$ \LRD, through which integrable representations, attached
to Young tableaux of at most $n$ rows and $k$ columns ($\widehat{sl(n)}_k$)
are mapped onto the dual (transposed) 
ones with at most $k$ rows and $n$ columns
($\widehat{sl(k)}_n$).  
A direct way to understand this duality, is
provided by the standard formulation \DJ\ \MUR\
of the basis 2, namely by
the use of a mapping between the basis 2 elements and 
the standard Young tableaux which have the shape of a rectangle of
$N$ rows by $k$ columns (basis of ${\cal I}_{Nk}^{(N)}(\beta)$) sent by
transposition to the standard Young tableaux having the shape of a rectangle
of $k$ rows by $N$ columns (basis of ${\cal I}_{Nk}^{(k)}(\beta)$).

A standard Young tableau of given shape $S$, $S$ a Young tableau of
$M$ boxes
(i.e. an arrangement of say $r$ rows of repectively $l_1$, $l_2$,...,$l_r$ 
square boxes, with $l_1\geq l_2\geq ...\geq l_r\geq 1$ and 
$l_1+l_2+...+l_r=M$),
consists of the tableau $S$, together with 
a labelling (marking)
of the boxes of $S$, using exactly once each of the integers 
$1,2,...,M$, and such that the labels are strictly increasing along
the rows (from left to right) and along the columns (from top to bottom).

In the particular case of a rectangular shape $S$ with $r=N$,
$l_1=l_2=...=l_N=k$, the set $S_{N,k}$ of the
corresponding standard tableaux is 
in bijection with the set of $SU(N)$
walk diagrams of order $Nk$.  
Indeed let us define the map 
$f\ :\ S_{N,k}\to W_{Nk}^{(N)}$, by sending any standard tableau
with $N$ rows and $k$ columns to the walk with successive
steps $v_i$,
$i=1,2,...,Nk$ defined by
\eqn\defvi{ v_i~=~ \epsilon_{r(i)} }
where $r(i)$ denotes the number of the row of the box marked $i$
in the standard tableau. For instance, the tableau whose marks
are entered by successive columns $(1,2,...,N)$, 
$(N+1,N+2,...,2N)$, ... $((k-1)N,(k-1)N+1,...,kN)$
is sent to the fundamental walk 
$a_0^{(N)}$, whereas the tableau whose marks are entered by successive 
lines $(1,2,...,k)$, $(k+1,k+2,...,2k)$, ..., 
$((N-1)k+1,(N-1)k+2,...,Nk)$
is sent to the maximal one $a_{max}^{(N)}$.

The map $f$ is clearly invertible, as 
we may fill the rectangular shape as we move along any walk $a$,
the $i$-th mark corresponding to the $i$-th step,
say $v_i=\epsilon_j$, and being
made in the leftmost available (unmarked) box of the $j$-th row,
thus filling eventually the whole
tableau, as there is an equal total number $k$ of steps of each kind 
$\epsilon_1$, ..., $\epsilon_N$.

The process of box addition at position $i$ on $a$
is interpreted in the standard tableau $f^{-1}(a)$ 
as the interchange of the
markings $i$ and $i+1$ if $i+1$ is in a strictly earlier row than $i$
(with $r(i+1)<r(i)$), this being
only possible if the ordering of rows
and columns is preserved by the interchange: 
this corresponds exactly to the situation
where the original walk has minimum at position $i$.

Now we see that the duality map $\delta$ has the simple 
interpretation as transposition, namely interchange of rows and
columns, in the standard tableau picture, namely
$f^{-1}(\delta(a))=f^{-1}(a)^t$, for all $a\in W_{NK}^{(N)}$.
Hence the map $f^{-1}\circ \delta \circ f$ is nothing but
the tableau transposition, which maps $S_{N,k} \to S_{k,N}$.
The dual correspondence \recdelta\
between box additions and subtractions 
becomes clear with the above interpretation: the
interchange between
the marks $i$ and $i+1$ has the effect of a box addition on  
a standard tableau iff it has the effect of a box subtraction on the 
transposed tableau.

\newsec{Hecke determinants}

In this section, we present determinant formulae for
the natural generalization of meander determinants to the whole
$SU(N)$ quotient $H_{n}^{(N)}(\beta)$ of the Hecke algebra.
This provides yet another direction of generalization of meanders.

\subsec{Bases of the Hecke algebra}
\par
The standard basis \DJ\ \CHE\ \MUR\ of the Hecke algebra 
$H_n(\beta)$ is indexed
by pairs $(s_1,s_2)$ of standard tableaux of $n$ boxes with the
same shape $S$, where $S$ describes the set of Young tableaux of $n$
boxes.  As already mentioned, this basis uses the description of the
Hecke algebra \hec\ in terms of the generators $T_i$ \genet.
The restriction of this basis to the quotient $H_n^{(N)}(\beta)$
is simply obtained by restricting the shapes $S$ to the tableaux
with at most $N$ rows.

Let us present now a slightly different basis of $H_n^{(N)}(\beta)$, 
which we call
basis 1 by analogy with the previous sections. This basis 1
will be indexed by pairs of {\it open} walk diagrams, rather than
standard tableaux; the
two objects are however in one-to-one correspondence. 

For any given
weight $\Lambda\in P_+$,
an {\it open} walk diagram of order $n$ ending at $\Lambda$
is a path of $n$ steps on $\Pi_+$, starting at the origin $\rho$
and ending at $\Lambda$. In particular, we must have $n-\sum i(\lambda_i
-1)=0$ mod $N$, if $\Lambda=(\lambda_1,...,\lambda_{N-1})$.
Let us denote by $W_{\Lambda}^{n}$ the set of open $SU(N)$
walk diagrams of order $n$, ending at $\Lambda$. 
Writing 
\eqn\lamwri{ \eqalign{
\Lambda~&=~ \rho+ \sum_{i=1}^N l_i \epsilon_i \cr
n~&=~\sum_{i=1}^N l_i \cr} }
easily inverted into $l_i=(\Lambda-\rho)\cdot \epsilon_i +n/N$ 
(see \inverty),
we may identify each walk diagram $a\in W_\Lambda^n$ with a standard
tableau whose shape is the Young tableau with $l_i$ boxes in the
$i$-th row, $1\leq i\leq N$. Indeed, the marking of the boxes corresponding
to $a$ is performed as one moves along the path; say when the $i$-th step 
is made, with $v_i=\epsilon_j$, we mark  with the integer $i$
the leftmost available (non-marked) box in the $j$-th row.
We have already 
computed in \coefn\ the number $C_{\Lambda}^{(n)}$
of open walk diagrams of order $n$ ending at $\Lambda$. 

The open walks of $W_\Lambda^n$ can be generated by box additions on
the fundamental one, denoted $a_0^{(n,\Lambda)}$, with steps
\eqn\stps{\eqalign{
v_{Ni+j}~&=~\epsilon_j \quad {\rm for}\ 
\left\{\matrix{ i=0,1,...,l_N-1 \cr  
j=1,2,...,N }\right. \cr
v_{Nl_N+(N-1)i+j}~&=~\epsilon_j \quad {\rm for}\  
\left\{\matrix{i=0,1,...,l_{N-1}-l_N-1\cr  
j=1,2,...,N-1\cr}\right. \cr
\ \ \cdots\ \  ~&~ \qquad \cdots \cr
v_{Nl_N+(N-1)(l_{N-1}-l_N)+...+2(l_2-l_3)+i+1}~&=~
\epsilon_1\quad {\rm for}\ i=0,1,...,l_1-l_2-1\cr}}
(with $l_i$ defined by \lamwri), 
which corresponds to entering the 
successive marks of the associated Young tableau by columns.
A box addition at position $i\in \{1,2,...,n-1\}$
on $a\in W_\Lambda^n$, denoted by $a \to a+\diamond_i$,
is defined in the same way as before (see Sect.4.2), and we still 
resolve the hexagon ambiguities by  forbiding the box additions of 
the form \foradbo.
This permits to construct all the walks of $W_\Lambda^n$ by successive
box additions on the fundamental $a_0^{(n,\Lambda)}$.

We are now ready to define the basis 1 elements of $H_n^{(N)}(\beta)$.
They are labelled by pairs $(a,b)$ of open walk diagrams belonging to
the same set $W_\Lambda^n$, where $\Lambda$ runs over $P_+$.
We start with the fundamental element
\eqn\fundonehe{\eqalign{
(a_0^{(n,\Lambda)},a_0^{(n,\Lambda)})_1~&=~
\prod_{i=0}^{l_N-1} E(e_{Ni+1},e_{Ni+2},...e_{Ni+N-1}) \cr
&\times \prod_{i=0}^{l_{N-1}-l_N-1} 
E(e_{Nl_N+(N-1)i+1},...,e_{Nl_N+(N-1)i+N-2})\cr
&\times \quad ... \cr
&\times \prod_{i=0}^{l_2-l_3-1} 
E(e_{Nl_N+(N-1)(l_{N-1}-l_N)+...+3(l_3-l_4)+2i+1}) \cr} }
where the antisymmetrizer $E$ is defined in \yonot, and related to
$Y$ through \propY\-\procons.
This product 
form corresponds to the column-preserving antisymmetrizer of \DJ\ 
\CHE.

The other basis 1 elements are defined recursively using box additions
on either walk diagram of the pair $(a,b)\in W_\Lambda^n$, namely
\eqn\boxhec{\eqalign{
(a+\diamond_i,b)_1~&=~ e_i (a,b)_1 \cr
(a,b+\diamond_j)_1~&=~ (a,b)_1 e_j \cr}}
We will call left (resp. right) box additions those pertaining to the
first (resp. second) line of \boxhec.
Note that the forbidden additions \foradbo\ make the box decompositions of
both $a$ and $b$ unique, and so is $(a,b)_1$.

For illustration, let us describe the basis 1 for $H_3^{(3)}$. 
There are three types of open walk diagrams of $3$ steps, namely
those which end at the $SU(3)$ weights $(1,1)$, $(2,2)$ or $(4,1)$,
with $|W_{(1,1)}^3|=1$, $|W_{(2,2)}^3|=2$ and $|W_{(4,1)}^3|=1$.
With $E(e_1,e_2)=Y(e_1,e_2)$ and $E(e_1)=e_1$, 
the basis 1 elements read respectively
\eqn\exbasi{\eqalign{
\left( \figbox{.8cm}{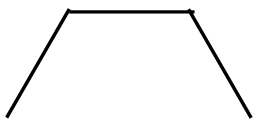},\figbox{.8cm}{l1.eps}\right)_1~&=~
Y(e_1,e_2)\cr
\left( \figbox{.8cm}{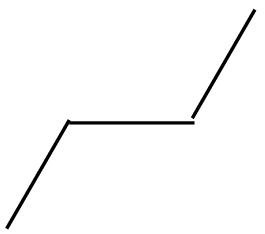},\figbox{.8cm}{l2.eps}\right)_1~&=~
e_1\cr   
\left( \figbox{.8cm}{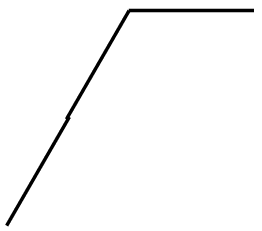},\figbox{.8cm}{l2.eps}\right)_1~&=~
e_2e_1\cr
\left( \figbox{.8cm}{l2.eps},\figbox{.8cm}{l3.eps}\right)_1~&=~
e_1e_2\cr
\left( \figbox{.8cm}{l3.eps},\figbox{.8cm}{l3.eps}\right)_1~&=~
e_2e_1e_2\cr
\left( \figbox{.8cm}{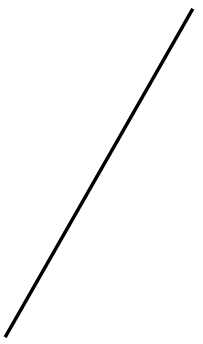},\figbox{.8cm}{l4.eps}\right)_1~&=~
1\cr}}
Note that the basis 1 for $H_3^{(2)}$ is simply obtained
by imposing
the vanishing of the antisymmetrizer of order $3$, namely by erasing the
first line in \exbasi: it consists of the five elements
$e_1$, $e_2e_1$, $e_1e_2$, $e_2e_1e_2=e_2$ and $1$.

As an immediate consequence, we get a formula for the dimension of
$H_n^{(N)}(\beta)$ as a vector space, namely
\eqn\dimhN{ \dim\big(H_n^{(N)}(\beta)\big)~=~\sum_{\Lambda\in P_+}
\big(C_\Lambda^{(n)}\big)^2  }
by enumerating all the pairs of open $SU(N)$ walk diagrams of order $n$.
The first few of these dimensions are displayed in Table VII.

$$\vbox{\font\bidon=cmr8 \def\bidondon{\bidon} \bidondon
\offinterlineskip
\halign{\tv \quad # \tv &
\hfill \ # & \hfill # & \hfill # &  \hfill #
& \hfill # &  \hfill #
& \hfill # &  \hfill #
& \hfill # &  \hfill #
\tv \cr
\noalign{\hrule}
\tvi  $\scriptstyle N\backslash n$& 1 \hfill & 2 \hfill
& 3 \hfill & 4 \hfill & 5 \hfill
& 6 \hfill 
& 7 \hfill 
& 8 \hfill 
& 9 \hfill 
& 10 \hfill 
\cr
\noalign{\hrule}
\tvi 2 & 1 & 2 & 5 & 14 & 42 & 132 & 429 & 1430 & 
4862 & 16796 \cr
\tvi 3 & 1 & 2 & 6 & 23 & 103 & 513 & 2761 & 15767 & 
94359 & 586590 \cr
\tvi 4 & 1 & 2 & 6 & 24 & 119 & 694 & 4582 & 33324 & 
261808 & 2190688 \cr
\tvi 5 & 1 & 2 & 6 & 24 & 120 & 719 & 5003 & 39429 & 
344837 & 3291590 \cr
\tvi 6 & 1 & 2 & 6 & 24  & 120 & 720 & 5039 & 40270 & 
361302 & 3587916 \cr 
\noalign{\hrule}
}}$$
\leg{{\bf Table VII:} The dimensions $\dim\big(H_n^{(N)}(\beta)\big)$
of the $SU(N)$ quotients of the Hecke algebra $H_n(\beta)$, with
$2\leq N\leq 6$ and $1\leq n\leq 10$.}

The Gram matrix ${\cal H}_n^{(N)}(\beta)$ of the basis 1 of 
$H_n^{(N)}(\beta)$ reads
\eqn\grah{ \big[ {\cal H}_n^{(N)}(\beta)\big]_{(a,b),(c,d)}~=~
\big( (a,b)_1,(c,d)_1\big)}
for $(a,b)$ and $(c,d)\in \cup_{\Lambda\in P_+}(W_\Lambda^n)^2$.
We wish to compute the determinant of this matrix exactly.
In the case $n=3=N$ of \exbasi, this Gram matrix reads
\eqn\grahec{ {\cal H}_3^{(3)}(\beta)~=~ (\b^2-1)
\pmatrix{\b^2(\b^2-1) & \b^2 & \b^3 & \b^3 & \b^4 & \b \cr
\b^2 & \b^2 (\b^2-1) & \b^3 & \b^3 & 2\b^2 & \b(\b^2-1) \cr
\b^3 & \b^3 & \b^4 & 2 \b^2 & 2 \b^3 & \b^2 \cr
\b^3 & \b^3 & 2 \b^2 & \b^4 & 2 \b^3 & \b^2 \cr
\b^4 & 2 \b^2 & 2 \b^3 & 2 \b^3 & 2\b^4 & \b^3 \cr
\b & \b(\b^2-1) & \b^2 & \b^2 & \b^3 & (\b^2-1)^2 \cr}}

The semi-normal basis of the Hecke algebra presented in \DJ\ and \MUR\
restricts up to normalization factors 
to the orthonormal basis 2 w.r.t. the
scalar product $(\ ,\ )$.  In our language, the basis 2 is
constructed as follows.
For each $\Lambda\in P_+$, we introduce the fundamental  
element
\eqn\funtwoH{ 
(a_0^{(n,\Lambda)},a_0^{(n,\Lambda)})_2~=~
\bigg(\prod_{{\rm all} \ {\rm steps}\atop
v\ {\rm of} \ a_0^{(n,\Lambda)} } w(v)\bigg)\   
g_\Lambda \ E_\Lambda(e_1,e_2,...,e_{n-1}) }
where, $g_\Lambda$ is a normalization constant
and $E_\Lambda$ are the orthogonal
idempotents of the semi-normal basis \DJ\ \MUR,
defined as follows, in terms of the Murphy operators $L_m$ \murop. 
We have
\eqn\defidem{
E_\Lambda~=~ \prod_{m=2}^n \prod_{-m<p<m\atop
p\neq 0 \ {\rm if}\ m=2,3} 
{L_p-[p]_q \over [r_\Lambda(m)]_q-[p]_q} }
where $[p]_q=1+q+q^2+...+q^{p-1}$ 
and $[-p]_q=-(q^{-1}+q^{-2}+...+q^{-p})$ for $p>0$, 
and $[0]_q=0$, and 
$r_\Lambda(m)=j-i$ if
\eqn\marko{ l_1+l_2+...+l_{i-1}< m\leq l_1+l_2+...+l_i }
for $l_i$ as in \lamwri, 
and $j=m-(l_1+...+l_{i-1})$. 
In the standard tableau formulation of $\Lambda$ (with marks entered
by successive columns), 
$i$ and $j$ are respectively
the numbers of the row and column in which the mark $m$ occurs. 
Moreover,
in \funtwoH, the normalization constant $g_\Lambda$ will
eventually ensure
that the basis 2 element has norm 1. 
Let us comment briefly on this normalization now.

Note first the existence of the ``inclusion"
order on $W_\Lambda^n$, which we now denote by $a\leq b$ iff 
$a\subset b$, namely if
$b$ can be obtained from $a$ by some box additions.
Moreover, this order is extended to all open walk diagrams
by deciding that $W_{\Lambda}^n\leq W_{\Lambda'}^n$ 
(we also write $\Lambda\leq\Lambda'$)
iff the 
weight $\Lambda'$ can be obtained from $\Lambda$ by successive ``pushes"
$p_{i,j}$, $N\geq i>j\geq 1$, defined as 
\eqn\lampush{
p_{i,j}(\Lambda)~=~\Lambda+\epsilon_i-\epsilon_j}
allowed only if the result is still in $P_+$.  
In the Young tableau formulation
of $\Lambda$, this amounts to ``pushing" the rightmost box in the $j$-th
row to the $i$-th row, and is allowed only if the result is still 
a Young tableau.
This gives an order $\leq$ on all open walk diagrams.
The change of basis 1 $\to$ 2 will be 
triangular with respect to $\leq$, 
namely 
\eqn\chbas{ (a,b)_2~=~\sum_{c \leq a\ , \ d \leq b} P_{(a,b),(c,d)} 
(c,d)_1 }
The normalization $g_\Lambda$ is chosen so that
\eqn\norel{ g_\Lambda E_\Lambda~=~(a_0^{(n,\Lambda)},a_0^{(n,\Lambda)})_1
+\sum_{\Lambda'\leq \Lambda\atop
\Lambda'\neq \Lambda} \sum_{a,b \in W_{\Lambda'}^n} 
\nu_{\Lambda,\Lambda'}^{(a,b)}
\ (a,b)_1 }
for some coefficients $\nu_{\Lambda,\Lambda'}^{(a,b)}$. The condition
\norel\ will enable us to orthogonalize the Gram matrix 
${\cal H}_n^{(N)}(\beta)$ by the Gram-Schmidt procedure, through a
triangular redefinition of its lines and columns. 
The value of $g_\Lambda$ can be found for instance in \MUR\ and reads 
\eqn\valglam{ g_\Lambda~=~(\alpha_N)^{-l_N}
(\alpha_{N-1})^{l_{N}-l_{N-1}} ... (\alpha_2)^{l_3 -l_2} }
in terms of the numbers $\alpha_j$ \procons\ and the integers
$l_j$ \lamwri.

The other basis 2 elements are obtained by (left and right) box
additions on the fundamental elements \funtwoH, according to the
following recursions
\eqn\recutwoH{\eqalign{
(a+\diamond_{i,m},b)_2~&=~\sqrt{\mu_{m+1}\over \mu_m}(e_i-\mu_m)
(a,b)_2\cr
(a,b+\diamond_{j,l})_2~&=~\sqrt{\mu_{l+1}\over \mu_l} (a,b)_2
(e_j-\mu_l)\cr}}
where $m$ and $l$ denote the heights of the box additions \highN.
The normalization of the basis 2 elements to unity
is a consequence of their orthogonality,
in the same way as before (see \simuop).

In the case of $H_3^{(3)}$, the normalized idempotents
$g_\Lambda E_\Lambda$ read, in terms of the $e_i$
\eqn\muridem{\eqalign{
g_{(1,1)}E_{(1,1)}~&=~Y(e_1,e_2)\cr
g_{(2,2)}E_{(2,2)}~&=~
e_1-\mu_1\mu_2 Y(e_1,e_2)~=~\mu_2 e_1 (1-\mu_1 e_2) e_1\cr
g_{(4,1)}E_{(4,1)}~&=~
(1-\mu_1 e_1)(1-\mu_2 e_2)(1-\mu_1 e_1)\cr}}
and we have the basis 2 elements 
\eqn\exbasi{\eqalign{
\left( \figbox{.8cm}{l1.eps},\figbox{.8cm}{l1.eps}\right)_2~&=~
\mu_1^2\mu_2 g_{(1,1)}E_{(1,1)} \cr
\left( \figbox{.8cm}{l2.eps},\figbox{.8cm}{l2.eps}\right)_2~&=~
\mu_1^2 \mu_2^{1/2} \mu_3^{1/2} g_{(2,2)}E_{(2,2)} \cr   
\left( \figbox{.8cm}{l3.eps},\figbox{.8cm}{l2.eps}\right)_2~&=~
\mu_1^{3/2} \mu_2 \mu_3^{1/2} (e_2-\mu_1) g_{(2,2)}E_{(2,2)} \cr
\left( \figbox{.8cm}{l2.eps},\figbox{.8cm}{l3.eps}\right)_2~&=~
\mu_1^{3/2} \mu_2 \mu_3^{1/2} g_{(2,2)}E_{(2,2)} (e_2-\mu_1) \cr
\left( \figbox{.8cm}{l3.eps},\figbox{.8cm}{l3.eps}\right)_2~&=~
\mu_1 \mu_2^{3/2} \mu_3^{1/2} (e_2-\mu_1) g_{(2,2)}
E_{(2,2)} (e_2-\mu_1) \cr
\left( \figbox{.8cm}{l4.eps},\figbox{.8cm}{l4.eps}\right)_2~&=~
\mu_1^{1/2}\mu_2\mu_3 \mu_4^{1/2}
g_{(4,1)}E_{(4,1)} \cr}}
where we have applied the box addition rules \recutwoH.

\subsec{Hecke Determinants}
\par
In this section, we compute the determinant 
$\Theta_n^{(N)}(\beta)$ of the Gram matrix 
${\cal H}_n^{(N)}(\beta)$ \grah\ of the basis 1 of $H_n^{(N)}(\beta)$.
The result takes the form
\eqn\forthe{\Theta_n^{(N)}(\beta)~=~ 
\prod_{m=1}^{n+1} (U_m)^{t_{m,n}^{(N)}}}
where $t_{n,m}^{(N)}$ are integers derived below.

In view of eqs.\chbas\-\norel, we deduce that,
in terms of the matrix elements of $P$, the desired determinant
reads
\eqn\desideter{ \Theta_n^{(N)}(\beta)~=~ 
\prod_{\Lambda \in P_+} \prod_{a,b\in W_\Lambda^n} P_{(a,b),(a,b)}^{-2}}

The diagonal terms $\pi_{(a,b)}\equiv P_{(a,b),(a,b)}$
in \chbas\ satisfy the double recursion relation
\eqn\rediater{ \pi_{(a+\diamond_{i,m},b+\diamond_{j,\ell})}^2~=~
{\mu_{m+1} \mu_{\ell+1}\over \mu_m \mu_\ell} \pi_{(a,b)}^2 }
and we have the fundamental elements
\eqn\fundapiaa{\pi_{(a_0^{(n,\Lambda)},a_0^{(n,\Lambda)})}^2~=~
\left(\prod_{{\rm all}\ {\rm steps}\atop
v\ {\rm of}\ a_0^{(n,\Lambda)} } w(v) \right)^2}
It is easy to solve \rediater\-\fundapiaa\ as
\eqn\soltout{\eqalign{ 
P_{(a,b),(a,b)}^2~&=~\prod_{{\rm all}\ {\rm steps}\atop
v,v' \ {\rm of} \ a,b} w(v)w(v')\cr 
&=~ P_{a,a}^{-2} P_{b,b}^{-2}\cr}}
where, in the last line we have 
recognized the matrix elements of the change of
basis 1 $\to$ 2 for the $SU(N)$ case \dNm,
with $w$ as in \wN.

We are now ready to compute the determinant \desideter, by use of the
definition \wN. Assembling all the contributions pertaining to $\mu_m$,
we find
\eqn\detfind{ \Theta_n^{(N)}(\beta)~=~\prod_{m=1}^{n+1} 
(\mu_m)^{-\theta_{m,n}^{(N)}} }
where
\eqn\thetfunc{ \theta_{m,n}^{(N)}~=~\sum_{p=0}^{n-1}
\sum_{\Lambda'\in P_+}  \sum_{1 \leq i<l\leq N}
\sum_{\Lambda\in P_+\atop
\Lambda\cdot (\epsilon_l-\epsilon_i)=m}
C_{\Lambda}^{(p)} C_{\Lambda+\epsilon_i,\Lambda'}^{(n-p-1)} 
C_{\Lambda'}^{(n)} }
This summarizes all the possible occurrences of $\mu_m$ in \desideter.
We have denoted by $C_{\Lambda,\Lambda'}^{(r)}$ the number of paths 
of $r$ steps on $\Pi_+$, from $\Lambda$ to $\Lambda'$, which reads
\eqn\cco{ C_{\Lambda,\Lambda'}^{(r)}~=~ \sum_{\sigma \in S_N}
(-1)^{l(\sigma)} D_{\Lambda'-\sigma(\Lambda)}^{(r)} }
where the necessary reflections (additions/subtractions) have
been performed on the paths on $\Pi$ from $\Lambda$ to $\Lambda'$.
The combination of $C$'s in \thetfunc\ stands for the total number
of pairs $a,b\in W_{\Lambda'}^n$, with one specified  edge
$(\Lambda,\Lambda+\epsilon_i)$. The edge indeed
separates one of the walks $a$ or $b$ into
two parts: (i) the portion between $\rho$ and $\Lambda$, of length $p$
(a total of $C_\Lambda^{(p)}$ paths) 
(ii) the portion between $\Lambda+\epsilon_i$ 
and $\Lambda'$, of length $n-p-1$ (a total of $C_{\Lambda+\epsilon_i,
\Lambda'}^{(n-p-1)}$ paths). The extra factor accounts for the
$|W_{\Lambda'}^n|=C_{\Lambda'}^{(n)}$ possibilities for the other walk.

The desired formula \forthe\ then follows from the definition \nota, with 
$t_{m,n}^{(N)}=\theta_{m,n}^{(N)}-\theta_{m+1,n}^{(N)}$.
The first few numbers $t_{m,n}^{(N)}$
are listed in the case of $N=3$ in Table VIII.

$$\vbox{\font\bidon=cmr8 \def\bidondon{\bidon} \bidondon
\offinterlineskip
\halign{\tv \quad # \tv &
\hfill \ # & \hfill\  # & \hfill\ # &  \hfill\ #
& \hfill\ # 
&  \hfill\ # &  \hfill\ #  & \hfill\ #
&  \hfill\ # &  \hfill\ # \tv \cr
\noalign{\hrule}
\tvi  $\scriptstyle m\backslash n$& 1 \hfill & 2 \hfill
& 3 \hfill & 4 \hfill & 5 \hfill
& 6 \hfill & 7 \hfill & 8 \hfill  & 9 \hfill & 10 \hfill \cr
\noalign{\hrule}
\tvi 1 & 0 & 1 & 5 & 21 & 85  & 331 & 1155 & 2688  & -7872 &
-196425\cr
\tvi 2 & 1 & 2 & 6 & 26 & 136 & 774 & 4599 & 28080 & 174951 & 
1108158\cr
\tvi 3 &   & 1 & 5 & 22 & 102 & 521 & 2933 & 17872 & 115344  &
774396\cr
\tvi 4 &   &   & 1 & 10 & 69  & 424 & 2528 & 15184 & 93537  &
595602\cr
\tvi 5 &   &   &   & 1  & 17  & 171 & 1395 & 10305 & 72513 &
499291\cr
\tvi 6 &   &   &   &    & 1   & 26  & 358  & 3746  & 33889 &
281728\cr
\tvi 7 &   &   &   &    &     & 1   & 37   & 666   & 8666 &
94096\cr
\tvi 8 &   &   &   &    &     &     & 1    & 50    & 1137 &
17952\cr
\tvi 9 &   &   &   &    &     &     &      & 1     & 65 &
1819\cr
\tvi 10 &   &   &   &   &     &     &      &       & 1 &
82\cr
\tvi 11 &   &   &   &   &     &     &      &       & &
1\cr
\noalign{\hrule}
}}$$
\leg{{\bf Table VIII:} The powers $t_{m,n}^{(3)}$ of $U_m$ in
the Hecke meander determinant $\Theta_{n}^{(3)}(\beta)$, 
for $n=1,2,...,10$.
The determinant of the matrix ${\cal H}_3^{(3)}(\beta)$
of \grahec\ is read in the 
third column to be $\Theta_3^{(3)}(\beta)=U_1^5 U_2^6 U_3^5 U_4$.}

\newsec{Conclusion}

\subsec{Generalized meanders}

In this paper, we have investigated two possible 
directions of generalization of the notion of meander. 
The first direction, developed in Sects.3 and 4, defines
the $SU(N)$ meanders of order $Nn$
as pairs $(a,b)$ of $SU(N)$ walk diagrams of $Nn$ steps;
to these objects we have associated the quantity 
$\big((a)_1,(b)_1\big)$, namely the scalar product of the two 
corresponding basis 1 elements of the ideal 
${\cal I}_{Nn}^{(N)}(\beta)$. This quantity however 
has a simple combinatorial interpretation only in the $SU(2)$ 
case, where it relates directly to the number of connected components
of the meander (see eq.\basone).
Unfortunately, we have not yet been able to 
find a good combinatorial interpretation for $N\geq 3$,
such as formulations as (polymer or membrane) folding problems
for instance. We intend to return to this aspect
in a later publication.

The second possible direction, developed in Sect.5, would
rather define meanders as pairs of {\it couples} of open
$SU(N)$ walk diagrams of $n$ steps ending at some weight 
$\Lambda\in P_+$, the Weyl chamber of $sl(N)$.
Remarkably, the two pictures coincide for $N=2$, thanks to
the existence of an isomorphism between the left ideal 
${\cal I}_{2n}^{(2)}(\beta)=H_{2n}^{(2)}(\beta)e_1 e_3...e_{2n-1}$ 
and the Temperley-Lieb algebra $H_n^{(2)}(\beta)$. Schematically,
this is due to the two equivalent formulations of a walk diagram
of order $2n$ as a path $a\in W_{2n}^{2}$ 
from the origin $1$ to itself, or the pair
formed by its first and second halves (respectively a path of
$n$ steps say from the origin $1$ to the weight $m$, and from 
the weight $m$ to the origin $1$), which, up to reversal of the second
half, form a pair $(a',b')\in W_m^{n}$.
This breaks down for $N\geq 3$, as in the pair $(a',b')\in W_\Lambda^{n}$ 
the ``return path" $b'$ has to be described in the reverse order, from 
$\Lambda^t$ to $\rho^t=\rho$, and we cannot identify the pair with 
a walk diagram of $2n$ steps, starting and ending at the origin $\rho$
(there will be in general a necessary jump from $\Lambda$ to $\Lambda^t$,
or alternatively a necessary reversal of all directions on $\Pi_+$
for the return path).

The results of Sects.3,4 however seem to suggest that the first 
generalization is the good one, as the results for the meander
determinants take simple generalized
forms, which we could not find for the second generalization of Sect.5.

\subsec{Generalized semi-meanders}

The study of the whole Hecke quotient $H_n^{(N)}(\beta)$ has the 
advantage of offering a better framework to generalize the notion
of semi-meander, introduced in \DGG\ \DGGB\ \DGGT.
The original (multi-component)
semi-meander problem is that of enumerating 
the topologically inequivalent configurations of a (several) 
nonselfintersecting loop(s), crossing a half-line through $n$ 
given points. Any such configuration is called a (multi-component) 
semi-meander of order $n$.
In comparison with the meander case, the novelty is that loops can freely
wind around the origin of the half-line. The winding number is then defined
as the minimal number of intersections which would be created by replacing
the half-line with a line, plus one (this one was not added in the
definition of the winding used in \DGG, it permits however to present 
a more unified notion when discussing generalizations). 
Considering that the line 
separates the semi-meander
into an upper and lower ``open" arch configurations
it is easy to interpret any semi-meander with winding $m$ as a pair of
open walk diagrams of order $n$ ending at the weight $m$, or
equivalently with a basis 1 element of the Temperley-Lieb algebra
$H_n^{(2)}(\beta)$, corresponding to a walk diagram 
of order $2n$ with middle weight $\lambda_n=m$.

This suggests the following generalization of semi-meanders.
We will call $SU(N)$ semi-meander of order $n$
with winding $\Lambda\in P_+$ 
any pair of open walk diagrams $a,b\in W_\Lambda^{n}$.
The semi-meander matrix for order $n$ and winding $\Lambda$
is then defined, using the basis 1 of 
$H_n^{(N)}(\beta)$ described in Sect.5, as the Gram matrix with entries
\eqn\grasem{\big[ {\cal H}_\Lambda^{(n)}(\beta)\big]_{a,b}~=~
{\rm Tr}\big( (a,b)_1 \big) \quad a,b \in W_\Lambda^{n} }
Note that when $\Lambda=\rho$ (only possible if the order $n$ is
of the form $Nk$ for some integer $k$), 
this matrix is identical to the $SU(N)$
meander matrix \graN, which suggests to interpret a semi-meander
with winding $\rho$ as a meander. This is of course a consequence
of the identification $W_\rho^{Nk}\simeq W_{Nk}^{(N)}$ between
the open walk diagrams ending at the origin and the walk diagrams of
same order.

In \MOI, we have derived a simple formula for the determinant of these
matrices, when $N=2$. The strategy used was again a direct 
orthogonalization of the basis 1 of the corresponding vector space,
leading to a basis 2' strictly distinct from the restriction of the
basis 2 of $H_n^{(2)}(\beta)$. In the general $N\geq 3$
case, we expect the determinant
of \grasem\ to still be given by some simple product formula involving
the Chebishev polynomials \ceby.

\par
\vskip 2.cm

\noindent{\bf Acknowledgements}

We are thankful to I. Cherednik and T. Nakanishi for valuable 
dicussions.

\listrefs
\bye